\documentclass[aps, 10pt, nofootinbib,twocolumn, pre, superscriptaddress]{revtex4-1}

\pdfoutput=1   
 
\usepackage{caption}
\usepackage[T1]{fontenc} 
\usepackage[makeroom]{cancel}
\usepackage{physics}
\usepackage{empheq}
\usepackage{graphicx}
\usepackage{subcaption}
\usepackage{tabularx}
\usepackage{natbib}

\usepackage[
colorlinks=true,        
citecolor=blue,         
linkcolor=blue,         
urlcolor=blue           
]{hyperref}             
\usepackage{xcolor}    

\usepackage{booktabs}
\usepackage{latexsym}

\usepackage{mathrsfs}
\usepackage{amsmath}
\usepackage{amssymb}

\usepackage{enumitem}

\usepackage{lmodern}
\usepackage[T1]{fontenc}
\usepackage{txfonts}
\usepackage{color}

\usepackage{float}
\usepackage{placeins}

\usepackage{ragged2e}
\usepackage{varioref}
\colorlet{shadecolor}{lightgray}
\usepackage{hyperref}
\usepackage{mathtools}
\usepackage{parskip}

\usepackage{capt-of}
\usepackage{enumitem}

\usepackage{graphicx}
\usepackage{cleveref}

\newcommand{\comment}[1]{}

\newcommand {\mathsym}[1]{{}}
\newcommand {\unicode}[1]{{}}

\newcommand{\nn}{\nonumber}

\tabularnewline

\def \be {\begin{equation}}
\def \ee {\end{equation}}
\def \bea {\begin{eqnarray}}
\def \eea {\end{eqnarray}}
\newcommand{\eq}[1]{(\ref{#1})}

\begin{document}
\title{Work Statistics via Real-Time Effective Field Theory: Application to Work Extraction from Thermal Bath with Qubit Coupling}

\author{Jhh-Jing Hong}
\email{hongjhhjing06132061@gmail.com}
\affiliation{Department of Physics, National Taiwan Normal University, Taipei, Taiwan 11676}

\author{Feng-Li Lin}
\email{fengli.lin@gmail.com, Corresponding author}
\affiliation{Department of Physics, National Taiwan Normal University, Taipei, Taiwan 11676}

\date{\today}
\begin{abstract}
Quantum thermal states are known to be passive, as required by the second law of thermodynamics. This paper investigates the potential for work extraction by coupling a thermal bath to a qubit of either spin, fermionic, or topological type, which acts as a quantum thermal state at different temperatures. The amount of work extraction is derived from the work statistics under a cyclic nonequilibrium process. Although the work statistics of many-body systems are known to be challenging to compute explicitly, we propose an effective field theory approach to address this problem by assuming that the externally driven source couples to a specific quasiparticle operator in the thermal state. We show that the work statistics can be expressed succinctly in terms of this quasiparticle's thermal spectral function. We obtain the non-perturbative work distribution function (WDF) for a pure thermal bath in the absence of qubit coupling. With qubit coupling, we obtain the second-order WDF, from which the physical regime for work extraction can be precisely pinned down to help devise quantum heat engines or refrigerators. Their efficiency or coefficient of performance (COP) can be inferred from the combination of the fluctuation theorem and the first law, and we find that the spin/topological qubit-bath system generally yields a far better heat engine/refrigerator than the other two alternatives due to the underlying quantum statistics. 

\end{abstract}

\maketitle

\section { Introduction} 
\begin{figure}[htb!]
    \centerline{\includegraphics[width=0.95\linewidth]{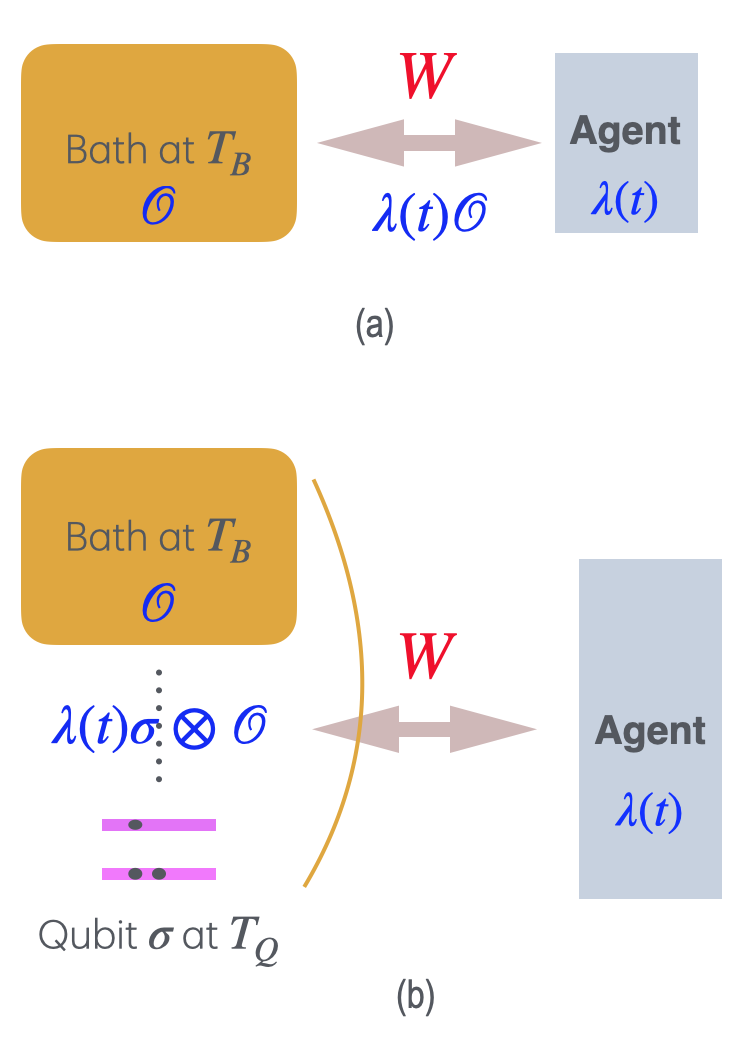}}
    \caption{Our effective field theory setups for considering the statistics of fluctuating work $W$: (a) An external agent drives the thermal bath away from thermal equilibrium by exciting a quasiparticle operator ${\cal O}$ with a time-varying coupling function $\lambda(t)$, which takes a cyclic profile. (b) Add an equilibrium-breaking qubit $\sigma$; its temperature $T_Q$ is different from $T_B$ of the thermal bath. The external agent then excites the qubit and the quasiparticle operator ${\cal O}$ together with the cyclic time-varying coupling $\lambda(t)$. This two-bath setup can act as a heat engine or a refrigerator operated by the external agent.
    }
\label{fig_scheme}
\end{figure}

Heat engines can extract work from the energy flow between two thermal baths in a cyclic irreversible process. Thus, no work extraction is possible with a single thermal bath, i.e., thermal states are passive \cite{Lenard1978ThermodynamicalPO, pusz1978passive, PhysRevE.91.052133}, as implied by the second law of thermodynamics. This was usually formulated in the thermodynamic limit. However, work is not a state variable; it fluctuates significantly for a small system and shall be characterized by work statistics. One can generalize the above second-law statement in terms of work statistics by the Jarzynski-Crooks fluctuation theorem (JC-FT)  for the cyclic processes \cite{jarzynski1997nonequilibrium, crooks2000path, crooks1999entropy, tasaki2000jarzynski}
\be\label{JC-FT}
\frac{P(-W)}{P(W)}=e^{-\beta W}  \quad {\rm or} \quad 1= \int_{-\infty}^{\infty} e^{-\beta W} P(W) dW := \overline{e^{-\beta W}}\;. 
\ee
Here, $P(W)$ is the work distribution function (WDF) of work $W$ done on a thermal state of inverse temperature $\beta$ in a cyclic irreversible process, and $P(-W)$ represents the WDF for the corresponding backward process. This relation gives a fine-grained definition of irreversibility, i.e., $P(-W)/P(W) < 1$ for $W > 0$. It can be verified experimentally by some driven protocol with a time-dependent source coupled to a classical thermal bath \cite{PMID:12052949, PhysRevLett.109.180601}, to a quantum thermal state \cite{PhysRevLett.113.140601, PhysRevLett.99.068101, PhysRevA.99.052508}, and to the quantum systems implemented or simulated on a quantum computer \cite{solfanelli2021experimental, hahn2023quantum, BassmanOftelie:2024rsn}. By Jensen's inequality, from \eq{JC-FT} we have
\be\label{W_2nd_law}
W_{\rm ext}=-\overline{W}:=-\int_{-\infty}^{\infty} W \; P(W) \; dW < 0\;,
\ee
where $W_{\rm ext}$ is the work extraction by an external agent acting on the thermal state by a driven protocol, see Fig. \ref{fig_scheme}(a). This reflects the fact that work extraction from a thermal state is impossible. 

The WDF is generally difficult to evaluate directly for many-body systems, such as quantum field theories, due to the complication of non-equilibrium many-body dynamics. See \cite{silva2008statistics, paraan2009quantum,  PhysRevB.93.104302, PhysRevB.100.035124, Ortega:2019etm, PhysRevE.100.052136} of the WDF for some exactly solvable many-body systems. On the other hand, its Fourier transform, the so-called work characteristic function (WCF), 
\be\label{chi_def0_m}
\chi(v)=\int_{-\infty}^{\infty} dW e^{i v W} P(W)\;,
\ee
can be written as the real-time correlation function \cite{talkner2007fluctuation1} defined on extended Schwinger-Keldysh contour \cite{fei2020nonequilibrium, Aron:2017spi, PhysRevE.100.062107}, so that it is more suitable for the effective field theory (EFT) approach adopted in this paper. In terms of $\chi(v)$ defined by \eq{chi_def0_m}, the JC-FT can be reforumlated as 
\be\label{JC-FT_i}
\chi(i\beta)=1\;.
\ee
Moreover, the WCF is the generating function of the work statistics, from which we have
\be\label{W_ext}
W_{\rm ext} := i \lim_{v\rightarrow 0} {\partial \chi(v)\over \partial v}\;.
\ee

Despite the passive nature of the thermal bath, exploring the possibility of extracting work by slightly breaking its thermal equilibrium is interesting.  The simplest way is to add a small system, such as a qubit, with its temperature \footnote{The qubit can be in a thermal mixed state, with which a temperature is associated based on the relative level populations.  In this sense, there exists a second thermal bath that brings the ensemble of qubits to an equilibrium mixed state.} different from that of the thermal bath. In this case, the small system can act as a second bath, allowing an external agent to possibly extract work from the heat flow between them, as in a conventional heat engine.  This also means that the JC-FT formulated by \eq{JC-FT}, and its derived relations \eq{W_2nd_law} and \eq{JC-FT_i}, do not hold exactly. In this paper, we will study the work statistics of cyclic irreversible processes controlled by an external agent acting on a thermal bath or with an appended qubit in the EFT approach. The agent also collects the work extracted during a cyclic process. In this approach, we assume the agent can couple to a particular observable or quasiparticle excitation of the thermal bath, or to the combined system of the qubit and thermal bath, via a time-dependent coupling strength. Then, the WCF can be expressed in terms of the real-time correlation functions (or the spectral density) of the quasiparticle-like observable, thereby bypassing the difficulties encountered in directly solving non-equilibrium many-body dynamics. We summarize our EFT setups\footnote{Interestingly, this EFT setup is quite parallel with the post-Newtonian EFT approach proposed in \cite{Lin:2024bmg} for the study of reduced dynamics of the qubit-like Unruh-DeWitt (UDW) type detectors \cite{PhysRevD.14.870, DeWitt:1980hx} near a black hole as a thermal bath.} for studying work statistics in Fig. \ref{fig_scheme}.

Using the EFT approach, in this paper, we can show that the JC-FT for the pure thermal bath holds to all orders of the driven coupling, and so does the second law $W_{\rm ext}<0$. This shows the passivity of thermal states explicitly and is a nontrivial all-order generalization of the result of  \cite{Bartolotta:2017rth}, which considered the work statistics under a $\lambda(t) \phi^4$ driven protocol with $\phi$ a relativistic scalar field based on the perturbation method of scattering amplitude up to ${\cal O}(\lambda^2)$. As for the setup in Fig. \ref{fig_scheme}(b) with an equilibrium-breaking qubit attached to thermal bath, the resultant amount and the sign of the work extraction $W_{\rm ext}$ depends on (i) the source function of the driven protocol, (ii) the spectral density of the thermal operator, (iii) types of qubit, e.g., spin, fermion of Dirac or Majorana \cite{ho2014decoherence, liu2016decoherence} types as considered below, (iv) the spectral gap of the qubit Hamiltonian, and (v) the respective temperatures of thermal bath and qubit. We can determine the $W_{\rm ext}>0$ regime by systematically surveying the parameter space. This can help build a quantum engine with a single heat bath and a qubit.

The rest of the paper is organized as follows. In the next section, we extend the fluctuation theorem for the work statistics to that of entropy production for non-thermal states, and derive the efficiency/coefficient of performance of the heat engine/refrigerator for the setup Fig. \ref{fig_scheme}(b). 
In section \ref{sec:3}, we provide a brief review of work statistics in the real-time formalism and extend it to the EFT context. In section \ref{sec3}, we apply the EFT formalism to obtain the work statistics of the pure thermal bath, and then in section \ref{sec4} to the cases with qubit couplings to the thermal bath. In section \ref{sec5}, we present the numerical patterns of work extraction based on the analytical results in section \ref{sec4}. We then conclude this work in section \ref{sec6}. Technical details are relegated to the Appendices. We give a thorough review in Appendix \ref{app_A} on the real-time formalism of work statistics. In Appendix \ref{app_wick}, we review the generalized Wick's theorem and then apply it to derive the non-perturbative work statistics for pure thermal states. In Appendix \ref{app_C} and \ref{app_D}, we give the details of deriving work statistics for the spin qubit coupling and fermion/topological qubit coupling, respectively. Finally, we provide some supplementary plots in Appendix \ref{app_E}.

\section{Entropy production, work statistics, heat engine/refrigerator and all that}

Although the original consideration of the JC-FT and work statistics in \cite{jarzynski1997nonequilibrium, crooks2000path, crooks1999entropy, tasaki2000jarzynski} is only for the pure thermal states, it can be naturally extended to the diagonal states rather than the thermal states in the conventional two-point measurement (TPM) scheme, for example, the case as in Fig. \ref{fig_scheme}(b). Especially, the WDF and WCF can still be defined. We have illustrated this extension explicitly in Appendix \ref{app_A}. Moreover, the JC-FT \eq{JC-FT} no longer holds and should be modified into \cite{PhysRevLett.71.2401, PhysRevE.50.1645, PhysRevLett.74.2694, crooks1999entropy} 
\be\label{JC-FT_full}
\frac{P(-\sigma)}{P(\sigma)}=e^{-\sigma}  \quad {\rm or} \quad  \overline{e^{-\sigma}}=1\;. 
\ee
where $\sigma$ is the entropy production during the cyclic process. Thus, by Jensen's inequality, this implies the second law $\Delta S:=\overline{\sigma} \ge 0$, where $\Delta S$ is the average entropy change for one cycle of the cyclic process. Moreover, this fluctuation theorem \eq{JC-FT_full} for the entropy production should hold for generic states under either cyclic or non-cyclic irreversible processes, as shown in \cite{PhysRevA.77.034101, PhysRevX.6.041017, aaberg2018fully, PhysRevX.9.031029, Huang:2022tup}.

When considering the non-cyclic process for the work statistics of a single thermal bath, the fluctuation theorem also takes the form of \eq{JC-FT_full} with the first law relation $\sigma=\beta (W - \Delta F)$, where $\Delta F$ is the difference between the free energies of the final state and initial state. In contrast, $\Delta F=0$ for the cyclic processes, then $\Delta S= \beta W$ if there is only a single thermal bath. Otherwise, the first law relation should be replaced by the one with two thermal baths, as discussed below.

As mentioned, the setup Fig. \ref{fig_scheme}(b) can be viewed as a heat engine/refrigerator with two heat baths at temperatures $T_B$ and $T_Q$, respectively. Assume $Q_B$ and $Q_Q$ are, respectively, the heat flows into the bath and the qubit. Then, by the first law relations of the adiabatic processes\footnote{This is consistent with the EFT consideration by requiring the externally driven coupling for the cyclic processes to be adiabatically varying in time.}, one has
\be\label{thermo}
\Delta S= Q_B/T_B + Q_Q/T_Q \;, \qquad \overline{W} = Q_B+Q_Q\;. 
\ee
In our EFT approach, we cannot obtain $Q_B$ and $Q_Q$ directly. However, we can obtain $\Delta S$ from the fluctuation theorem and $\overline{W}$ from the WCF by \eq{chi_def0_m}, then we can solve \eq{thermo} to express $Q_{B,Q}$ in terms of $\Delta S$ and $\overline{W}$, i.e., $Q_B=-{T_B (T_Q \Delta S - \overline{W}) \over T_B- T_Q}$ and $Q_Q={T_Q(T_B \Delta S - \overline{W}) \over T_B - T_Q}$.

If the $W_{\rm ext}= -\overline{W} > 0$, the setup of Fig. \ref{fig_scheme}(b) functions as a heat engine with the efficiency 
\be\label{efficiency}
\eta := W_{\rm ext} /|Q_H| =  {1-r \over 1 + { T_L \Delta S \over W_{\rm ext}}}  \qquad {\rm with} \quad r:={T_L \over T_H} \;,
\ee
where $H$ ($L$) labels the quantities associated with one of the baths with higher (lower) temperature, namely, $T_H:={\rm max}(T_B, T_Q)$ and $T_L:={\rm min}(T_B, T_Q)$. Note that $0\le \eta \le 1-r$, where the upper bound is the Carnot efficiency and is reached when $\Delta S=0$. On the other hand, if $\overline{W}>0$, the same setup now functions as a refrigerator with its coefficient of performance (COP)  
\be\label{COP}
{\rm COP} := |Q_L|/\overline{W} =  {r \over 1- r} ( 1 - {T_H \Delta S \over \overline{W}})  \;.
\ee
For $\Delta S=0$, the above COP reduces to the ${r\over 1-r}$ of the Carnot refrigerator.

We will not introduce a work storage system, but will implicitly assume the external agent can carry out the task. This is similar to the original consideration in \cite{jarzynski1997nonequilibrium, crooks2000path, crooks1999entropy, tasaki2000jarzynski} \footnote{The external agent is either explicitly mentioned in \cite{tasaki2000jarzynski}, or implicitly introduced in \cite{jarzynski1997nonequilibrium, crooks2000path, crooks1999entropy} by emphasizing the work is done on the bath by evolving the external parameters. }, and see \cite{2016PhRvX...6d1017A, aaberg2018fully} for the consideration with an additional storage system.  We explicitly calculate in section \ref{sec5} the efficiency/COP of the heat engine/refrigerator of Fig. \ref{fig_scheme}(b) for the qubit-bath coupling in the framework of real-time EFT.  

If the thermal stats are replaced by non-diagonal ones, the TPM will then destroy the initial coherence, so that one cannot define the WDF or WCF to satisfy some general fluctuation theorem. Instead, the entire driven process should be treated as a quantum channel, and the reverse process for the generalized fluctuation theorem is the Petz recovery map \cite{Petz:1986tvy}, with the initial non-diagonal state serving as the reference state. The generalized fluctuation theorem will then be reformulated as constraints on the exchange of information and entropy production \cite{2016PhRvX...6d1017A, aaberg2018fully, kwon2019fluctuation, bai2024fully}. This, however, falls beyond the scope of our current study.

\bigskip

\section {Work statistics via real-time EFT}
\label{sec:3}
We will consider the work statistics for the following driven protocol described by the Hamiltonian 
\be\label{H_tot_m}
H(t) = H_0 + \lambda(t) V
\ee
where $H_0$ is the kinetic term of the system, and $V$ is the hermitian operator, e.g., an elementary or composite field, coupled to the externally driven source $\lambda(t)$. For real but nontrivial $\lambda(t)$, $H(t)$ is hermitian but non-conserving. In addition, the interaction Hamiltonian should take the generic form $\int d^3x \lambda(x^{\mu}) V(\vec{x})$, which indicates spatial dependence of the source function. For simplicity, however, we assume $\lambda(x^{\mu})$ to be either spatially uniform, i.e., $\lambda(x^{\mu})=\lambda(t)$ or ultralocal, i.e., $\lambda(t)\delta(\vec{x}- \vec{x}_0)$, so that the interaction Hamiltonian can reduce to the form in \eq{H_tot_m} with $V=\int d^3x V(\vec{x})$ or $V(\vec{x}_0)$, respectively. In this paper, the system we will consider is either a thermal bath or a qubit coupled to a specific thermal operator, made of a spin, a Dirac fermion, or two spatially separated Majorana zero modes.

\begin{figure}
\centering
\includegraphics[width=0.4\textwidth]{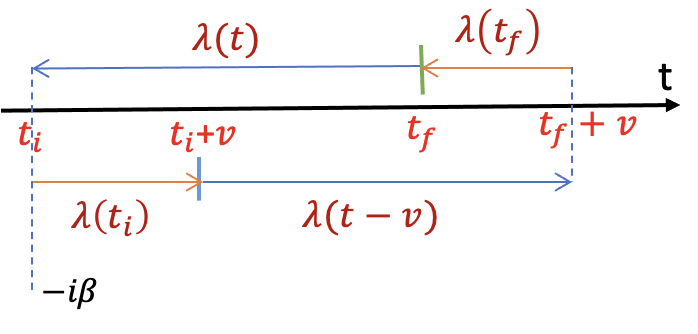}
\caption{\label{contour} The Schwinger-Keyldysh contour for defining the WCF $\chi(v)$ of \eq{chi_realtime_m} for the work statistics. The forward and backward contours are formally separated by the small artificial displacements below and above the real $t$ axis, respectively. We have included the finite-temperature path of the initial thermal state along the imaginary axis.}
\end{figure}

Evolving a system's initial state $\rho$ by $H(t)$ during the time interval $[t_i, t_f]$, the corresponding WCF can be expressed as the following real-time correlator, as firstly shown in \cite{talkner2007fluctuation1},
\be\label{chi_def_m}
\chi(v)={\rm Tr}\big[ U_S(t_f,t_i) e^{-i H(t_i) v} \; \rho \; U_S(t_i, t_f) e^{iH(t_f) v} \big]
\ee
where $U_S(t_2,t_1)$ is the evolution operator in the Schrödinger picture. To adopt the real-time formalism, we translate \eq{chi_def_m} from Schrödinger picture to interaction picture by expressing $U_S(t_2,t_1)$ as $e^{-i H_0 t_2}   {\cal T}_{\pm} \big[ e^{-i \int_{t_1}^{t_2} \lambda(t) V_I(t) dt} \big]  e^{i H_0 t_1}$ with $V_I(t):=e^{i H_0 t} V e^{-i H_0 t}$ and the choice of forward time-ordering ${\cal T}_+$ for $t_2\ge t_1$ or backward time-ordering ${\cal T}_-$, otherwise. The time orderings are defined by   ${\cal T}_+\big[A(t)B(t')\big] := \Theta(t-t') A(t)B(t') +  \Theta(t'-t) B(t') A(t)$ and ${\cal T}_-\big[A(t)B(t')\big] :=  \Theta(t'-t) A(t)B(t') +  \Theta(t-t')  B(t') A(t)$ for bosonic operators $A$ and $B$.  After some calculations, as reviewed in Appendix A, one can obtain $\chi(v)$ as the real-time correlator in the extended Keldysh contour of Fig.\ref{contour} as follows \cite{fei2020nonequilibrium}:
\be
\chi(v) = {\rm Tr}\Big\{{\cal T}_+ \big[ e^{-i \int_{-\infty}^{\infty} dt \; \lambda^+(t;v) V_I(t)} \big] \; \rho_I \; {\cal T}_- \big[ e^{i \int_{-\infty}^{\infty} dt \; \lambda^-(t;v) V_I(t)} \big]
 \Big\} \;. \label{chi_realtime_m}
\ee
where $\rho_I:=e^{i H_0 t_i} \rho e^{- i H_0 t_i}$ and 
\bea\label{lambda_p_m}
\lambda^+(t;v) &:=& \lambda_i \Theta(t;t_i,t_i+v)
 + \lambda(t-v) \Theta(t;t_i+v,t_f+v)\;, \\
\lambda^-(t;v) &:=& \lambda(t) \Theta(t;t_i,t_f) + \lambda_f \Theta(t;t_f,t_f+v) \;, \label{lambda_m_m}
\eea
where $\Theta(t;t_1,t_2)$ is  unity if $t\in [t_1,t_2]$, and vanishes, otherwise.

Even though \eq{chi_realtime_m} looks relatively compact and elegant, it is not easy to simplify it further due to the complications by (a) nontrivial  $\lambda^{\pm}(t;v)$ and (b) time orderings. We can resolve (a) by considering the cyclic processes with $t_{i,f}=\mp \infty$ and $\lambda(t_{i,f})=0$ so that \eq{lambda_p_m} and \eq{lambda_m_m} are reduced to the simpler forms:
\be\label{lambda_pm_simple}
\lambda^-(t;v)=\lambda(t)\;, \quad  \lambda^+(t;v)=\lambda(t-v)\;.
\ee
In this work, we will consider the Gaussian-driven source
\be\label{drive_s}
\lambda(t) = \lambda_0  (2 \pi)^{-1/4} e^{-t^2 \over 4 T_{\rm int}^2}
\ee
and its Fourier transform is
\be\label{lambdaW}
\tilde{\lambda}(\omega) = \lambda_0 (8\pi)^{1/4} T_{\rm int} e^{-\omega^2 T_{\rm int}^2}\;.
\ee
Note that $\int_{-\infty}^{\infty} dt \big\vert {\lambda(t) \over \lambda_0} \big\vert^2 = T_{\rm int}$ so that $T_{\rm int}$ can be thought as the effective interaction time. 
We also restrict $\rho$ to be diagonal in the eigenbasis of $H_0$, such as the thermal states. Then, the WCF of such cyclic processes still takes the form of \eq{chi_realtime_m} but with $\rho_I=\rho$ and $\lambda^{\pm}(t;v)$ given by \eq{lambda_pm_simple}.
On the other hand, the complication of time orderings prevents obtaining the results for all orders of $\lambda$ in general. Instead, we can carry out the WCF order by order in perturbation of $\lambda\ll 1$. Assuming ${\rm Tr}(\rho V)=0$, we consider WCF up to ${\cal O}(\lambda^2)$.

To proceed with further simplification, we also assume $\Tr(\rho V_I(t))=0$, i.e., no nonzero one-point function.
The result can be expressed in terms of real-time Green functions $G^{ab}(t-t')$ of $V_I(t)$ in the state $\rho$,
\be\label{chiv_2lambda_m}
\chi^{(2)}(v) = 1 - {i\over 2}\int_{-\infty}^{\infty} dt \int_{-\infty}^{\infty}dt' \sum_{a,b=\pm} \epsilon^{ab} \lambda^a(t;v) \lambda^b(t';v) \; G^{ab}(t-t') 
\ee
with $\epsilon^{ab}=1$ if $a=b$ and $\epsilon^{ab}=-1$ if $a\ne b$. The real-time Green functions of bosonic operator $V_I(t)$ are defined by  
\bea
&& i G^{++}(t,t'):=  \Tr \Big[ \rho {\cal T}_+ \big[V_I(t) V_I(t')\big] \Big]\;,  \\   
&&i G^{+-}(t,t'):=  \Tr \Big[ \rho V_I(t') V_I(t) \Big]\;, \label{real_G_1_0} \\
&& i G^{--}(t,t'):= \Tr \Big[ \rho {\cal T}_- \big[V_I(t) V_I(t')\big] \Big]\;, \\
&&i G^{-+}(t,t'):= \Tr \Big[ \rho V_I(t) V_I(t') \Big]\;. \label{real_G_2_0}
\eea
By construction, the real-time Green functions obey
\be\label{real_G_id}
G^{++}(t,t') + G^{--}(t,t') = G^{-+}(t,t') + G^{+-}(t,t') 
\ee
and  $G^{+-}(t,t') = G^{-+}(t',t)$. 
Moreover, the retarded Green function $G^R$ and the advanced  Green function $G^A$ are given by
\begin{align}
    \label{G_R0}
G^R(t,t') = & \, G^{++}(t,t') - G^{+-}(t,t') =G^{-+}(t,t') - G^{--}(t,t') \nonumber\\
= &\, -i \Theta(t-t') \Tr \Big( \rho [V_I(t),V_I(t')] \Big)\;,
\\
 G^A(t,t') = & \, G^{++}(t,t') - G^{-+}(t,t') = G^{+-}(t,t') - G^{--}(t,t') \nonumber\\ 
 = & \,i \Theta(t'-t) \Tr \Big( \rho [V_I(t),V_I(t')] \Big)\;. \label{G_sym}
\end{align}

We use the relations \eq{G_R0} to rewrite \eq{chiv_2lambda_m} into the following
\begin{widetext}
\begin{align}
    \chi^{(2)}(v) =& \, 1 -{1\over 2} \int_{-\infty}^{\infty} dt \int_{-\infty}^{\infty} dt' \; \Big\{ \big[ \lambda^+(t;v) \lambda^+(t';v) - \lambda^-(t;v) \lambda^-(t';v) \big] i G^R(t-t') \nn \\
& + \; \big[ \lambda^-(t;v) \lambda^-(t';v) - \lambda^-(t;v) \lambda^+(t';v) \big] i G^{-+}(t-t')  + \; \big[\lambda^+(t;v) \lambda^+(t';v) - \lambda^+(t;v) \lambda^-(t';v)\big] i G^{+-}(t-t') \Big\} \;.
\label{chiv_special}
\end{align}
\end{widetext}
Denoting the Fourier transform of a time domain function $f(t)$ by $\tilde{f}(\omega)$, and using the fact that $\int dt dt' \; A(t-a) B(t'- b) G(t-t') = \int {d\omega \over 2 \pi} \; e^{i \omega (a-b)} \tilde{A}(\omega) \tilde{B}(-\omega) \tilde{G}(\omega)$, and the fact that $\tilde{\lambda}(-\omega)=\tilde{\lambda}^*(\omega)$ because $\lambda(t)$ is real, the term involving $G^R$ disappears because $|\tilde{\lambda}^+(\omega)|^2 =|\tilde{\lambda}^-(\omega)|^2=|\tilde{\lambda}(\omega)|^2$. Thus, by \eq{lambda_pm_simple}, \eq{chiv_2lambda_m} can be expressed in the frequency domain and reduced to
\begin{align}
    \label{chiv_omega_1_m}
\chi^{(2)}(v)=\,1 -{1\over 2} \int_{\lambda} \Big\{ \big(1-e^{i\omega v} \big) i \tilde{G}^{-+}(\omega) + \big(1-e^{-i\omega v} \big)i \tilde{G}^{+-}(\omega) \Big\}\;.
\end{align}
In the above, we introduce the short-hand notation, which will be adopted from now on:
\be\label{short-h}
\int_{\lambda}:=\int_{-\infty}^{\infty} {d\omega \over 2\pi} \;  |\tilde{\lambda}(\omega)|^2\;.
\ee
Note that $\chi^{(2)}(0)=1$, consistent with $\int dW P(W)=1$.

Then, the WDF up to  ${\cal O}(\lambda^2)$ can be derived by inverse Fourier transform of \eq{chiv_omega_1_m},
\be\label{p_W_1_m}
P^{(2)}(W) = p^{(2)}_0 \delta(W) + {1\over 4\pi} |\tilde{\lambda}(W)|^2 \Big[  i\tilde{G}^{+-}(W) +  i\tilde{G}^{-+}(-W) \Big]
\ee
with
\be \label{m_p0}
p^{(2)}_0 := 1-{1\over 2} \int_{\lambda} \Big[ i\tilde{G}^{+-}(\omega) + i\tilde{G}^{-+}(\omega)  \Big] \;.
\ee
Although $P^{(2)}(W)$ is normalized by construction, it must also be positive definite. Thus, we may require $1 >P^{(2)}(W)\ge 0$ up to the order considered. By looking into \eq{p_W_1_m} and \eq{m_p0}, this implies
\be\label{constrain_P}
0<\int_{\lambda} \Big[ i \tilde{G}^{-+}(\omega) + i \tilde{G}^{+-}(\omega)  \Big] < 2
\ee
for $1> p^{(2)}_0 >0$. That is, we should choose the appropriate $\tilde{\lambda}(\omega)$, $\tilde{G}^{-+}(\omega)$, and $\tilde{G}^{+-}(\omega)$ so that \eq{constrain_P} is satisfied.

The generalization of JC-FT of \eq{JC-FT} up to ${\cal O}(\lambda^2)$ for state $\rho$ and interaction $V$ looks 
\be
{P^{(2)}(-W) \over P^{(2)}(W)}\Big\vert_{W\ne 0} = {\tilde{G}^{+-}(W) + \tilde{G}^{-+}(-W) \over \tilde{G}^{+-}(-W) + \tilde{G}^{-+}(W)}\;.
\label{JC-FT_g}
\ee
The right side is $\lambda$-independent. Moreover, we define the Pauli-Jordan's causal spectral function of the operator $V$ by 
\be
S^{V}(\omega):=i\tilde{G}^{-+}(\omega) -i\tilde{G}^{+-}(\omega)=\int_{-\infty}^{\infty} dt e^{i \omega t} {\rm Tr}\big(\rho [V_I(t), V_I(0)]\big)\;.
\ee
Then, the fine-grained irreversibility, i.e., $P(-W)<P(W)$, requires
\be\label{2nd_law_1}
S^{V}(\omega) > S^{V}(-\omega) \quad {\rm for} \quad \omega > 0\;.
\ee

We can also apply \eq{W_ext} to obtain work extraction up to ${\cal O}(\lambda^2)$,
\be\label{W_ext_f}
W^{(2)}_{\rm ext} = -{1\over 2} \int_{\lambda} \omega \; S^{V}(\omega)   \;.
\ee
Since $\lambda(t)$ is real, $\tilde{\lambda}(-\omega)=\tilde{\lambda}^*(\omega)$, hence $|\tilde{\lambda}(-\omega)|^2 = |\tilde{\lambda}(\omega)|^2$ and $W^{(2)}_{\rm ext}=-{1\over 2} \int_0^{\infty} {d\omega \over 2\pi} \; \omega \; |\tilde{\lambda}(\omega)|^2 \big[S^{V}(\omega) - S^{V}(-\omega) \big]$. Then, \eq{2nd_law_1} implies  $W^{(2)}_{\rm ext}\le 0$. Thus, microscopic irreversibility implies the passivity of the generic diagonal states. 

Finally, by the definition of WCF, we can arrive
\be
\overline{e^{-\beta W}} = \chi(i\beta)
\ee
for any parameter $\beta$. This equality should hold order by order in the expansion of $\lambda$. Thus, in the current consideration, we shall have 
\begin{align}
&\overline{e^{-\beta W}}\big\vert_{{\cal O}(\lambda^2)}=\chi^{(2)}(i\beta)
\nn\\
=&\, 1 -{1\over 2} \int_{\lambda} \big(1-e^{-\beta \omega} \big) \Big\{  i \tilde{G}^{-+}(\omega) - e^{\beta \omega}   i \tilde{G}^{+-}(\omega) \Big\} \;, \label{chi_beta2_G}
\end{align}
which is not equal to unity for general states. 
By Jensen's inequality $e^{-\beta \overline{W}} \le \overline{e^{-\beta W}}$,
\be\label{J_passive}
\overline{W}^{(2)} \ge -{1\over \beta} \ln \chi^{(2)}(i\beta)\;.
\ee
If we consider $\rho$ a thermal state of inverse temperature $\beta$, for which the Kubo-Martin-Schwinger (KMS) condition $\tilde{G}^{-+}(\omega) = e^{\beta \omega}\tilde{G}^{+-}(\omega)$ should hold such that the second term on the right side of \eq{chi_beta2_G} vanishes. Thus, under the cyclic process, the Jarzyski equality \eq{JC-FT_i} holds for the thermal state at ${\cal O}(\lambda^2)$. Later, we will show that it holds for all orders of $\lambda$. This then leads to $\overline{W}\ge 0$. This is the statement of the second law of thermodynamics. 
However, when coupling a qubit to a thermal state of inverse temperature $\beta$, we expect Jarzynski's equality to no longer hold. Thus, we can use $\chi^{(2)}(i\beta)$ to characterize the violation of Jarzynski equality up to ${\cal O}(\lambda^2)$. Thus, if $\chi^{(2)}(i\beta ) \ge 1$, the lower bound of $\overline{W}$ is negative; it is possible for the coupled qubit to extract work from the thermal environment. Otherwise, the lower bound is positive, and the total system remains passive. 

On the other hand, we can rewrite \eq{J_passive} into 
\be
\Delta S := \beta \overline{W}^{(2)} + \ln [\chi^{(2)}(i\beta)] \ge 0.
\ee
This is the second law statement implied by the general fluctuation theorem \eq{JC-FT_full} for entropy production. Based on this form $\Delta S$, we can evaluate the efficiency/COP for the setup of Fig. \ref{fig_scheme}(b) as a heat engine/refrigerator.

We have derived the WCF and WDF for the generic diagonal state. Below, we will consider work statistics first for the pure thermal bath and then with qubit coupling. 


\section { Work statistics of pure thermal bath}\label{sec3}

Consider the driven protocol \eq{H_tot_m} with $V=O$ a bosonic field operator of the thermal bath specified by thermal state $\rho=\rho_{\beta}$ of inverse temperature $\beta$, and $H_0=H_0^{\rm TB}$ the kinetic Hamiltonian of the thermal bath. The thermal real-time Green functions obey the KMS condition 
\be\label{KMS_m}
i \tilde{G}^{+-}(\omega) = n_{\rm BE}(\omega) S^O(\omega)\;, \quad  i\tilde{G}^{-+}(\omega) = e^{\beta \omega} n_{\rm BE}(\omega) S^O(\omega)  
\ee
with $n_{\rm BE}(\omega)=(e^{\beta\omega}-1)^{-1}$ and 
the Pauli-Jordan spectral function $S^O(\omega) := \int_{-\infty}^{\infty} dt e^{i \omega t} [O(t), O(0)] $ of bosonic quasiparticle operator $O$. The stability of the thermal state requires
\be\label{stability_m}
S^O(\omega) = 0, \quad {\rm for} \quad \omega <0\;.
\ee
The KMS condition implies the relation between the kernels of fluctuation and dissipation, known as the fluctuation-dissipation theorem:
\be\label{FDT_m}
S^O(\omega) = - 2{\rm Im}\tilde{G}^R(\omega)
\ee
with $\tilde{G}^R(\omega)$ the Fourier transform of the retarded Green function. For later convenience, we also introduce the Wightman spectral function $S_{\beta}(\omega):=\int_{-\infty}^{\infty} dt e^{i \omega t} {\rm Tr}[\rho_{\beta} O(t) O(0)]= i\tilde{G}^{-+}(\omega)$ for the thermal state $\rho_{\beta}$, which is related to Pauli-Jordan spectral function $S^O(\omega)$ by
\be
S_{\beta}(\omega) = e^{\beta \omega} n_{\rm BE}(\omega) S^O(\omega) \;.
\ee

\begin{figure}[htb!]
\centerline{\includegraphics[width=0.95\linewidth]{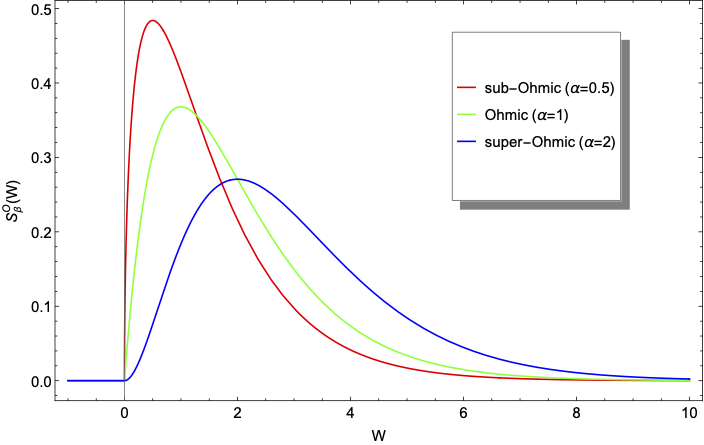}}
	\caption{Three Ohmic-type spectral density $S^O$ with $\alpha=0.5$ (red), $\alpha=1.$ (green) and $\alpha=2.$ (blue). We have set $\beta=1$, i.e., using $1/\beta$ as the unit for measuring $\omega$.}
\label{fig:ohmic}
\end{figure}

In this paper, we will consider the Ohmic-type spectral functions for either bosonic operator $O$ or fermionic operator $\Psi$:
\be\label{S_ohmic}
\quad S^{O \; {\rm or} \; \Psi}(\omega) = {2 L_c  (L_c \omega )^{\alpha} \over \Gamma\Big({1+\alpha \over 2}\Big)}  e^{-L_c^2 \omega^2 }\Theta(\omega) 
\ee
with $\alpha>0$ and $1/L_c$ the ultraviolet energy cutoff. The adiabatic condition  $T_{\rm int} \gg L_c$ shall be assumed to ensure the validity of the EFT approach. In all figures of this work, we fix $T_{\rm int}=100$ and $\lambda_0=0.01$ of the driven source \eq{lambdaW} and $L_c=1$ of the spectral function \eq{S_ohmic}. In Fig. \ref{fig:ohmic}, we show some examples of Ohmic-type spectral functions.

Plugging \eq{KMS_m} into \eq{chiv_omega_1_m} and \eq{p_W_1_m}, we can obtain the corresponding $\chi^{(2)}(v)$ and $P^{(2)}(W)$ as follows:
\be\label{chiv_omega_2}
\chi^{(2)}(v)=1-{1\over 2} \int_{\lambda} S^O_{\beta}(\omega) \Big\{ 1-e^{i\omega v} + e^{-\beta \omega} \big(1-e^{-i\omega v} \big) \Big\}\;,
\ee
\be\label{pW_1_m}
P^{(2)}(W) = p^{(2)}_0 \delta(W) + {1\over 4 \pi} |\tilde{\lambda}(W)|^2 \Big[ S^O_{\beta}(W) + e^{\beta W} S^O_{\beta}(-W) \Big]
\ee
with
\be
p^{(2)}_0 := 1-{1\over 2} \int_{\lambda} S^O_{\beta}(\omega) \Big(1+ e^{-\beta \omega}\Big)\;.
\ee
Along with \eq{stability_m}, it is straightforward to verify \eq{JC-FT} and \eq{JC-FT_i} explicitly, i.e., $\chi^{(2)}(i\beta)=1$, e.g.,
\bea
{P^{(2)}(-W) \over P^{(2)}(W)} &=& \begin{cases}
    1=e^{-\beta W}\;,  &  \quad \text{if } W=0\;, \\
    {S^O_{\beta}(-W) + e^{-\beta W} S^O_{\beta}(W) \over S^O_{\beta}(W) + e^{\beta W} S^O_{\beta}(-W) }=e^{-\beta W}\;,  & \quad \text{if } W\ne 0\;. \nn
  \end{cases}
  \label{Crooks relation thermal}
\eea
This reproduces the result of \cite{Bartolotta:2017rth} based on an alternative method.

Moreover, using \eq{FDT_m}, the work extraction obtained from \eq{W_ext_f} can be rewritten as 
\bea
W_{\rm ext} &=& \int_0^{\infty} {d\omega \over 2\pi} \; |\tilde{\lambda}(\omega)|^2 \omega \; {\rm Im}\tilde{G}^R(\omega)\;, \nn
\\
&=& -\int_{-\infty}^{\infty}dt \int_{-\infty}^{\infty}dt' {d \lambda(t) \over dt} G^R(t-t') \lambda(t')\;. \nn
\eea
This agrees with the $W_{\rm ext}=-\int_{-\infty}^{\infty} dt {d H(t) \over dt} = - \int_{-\infty}^{\infty} dt  {d \lambda(t) \over dt} \langle V_I(t) \rangle_{\lambda(t)}$ derived in \cite{biggs2024comparing} by using the linear response relation  $\langle V_I(t) \rangle_{\lambda(t)} = \int_{-\infty}^{\infty} dt' G^R(t-t') \lambda(t')$.

\begin{figure*}[htb!]
    \centerline{\includegraphics[width=1.05\linewidth]{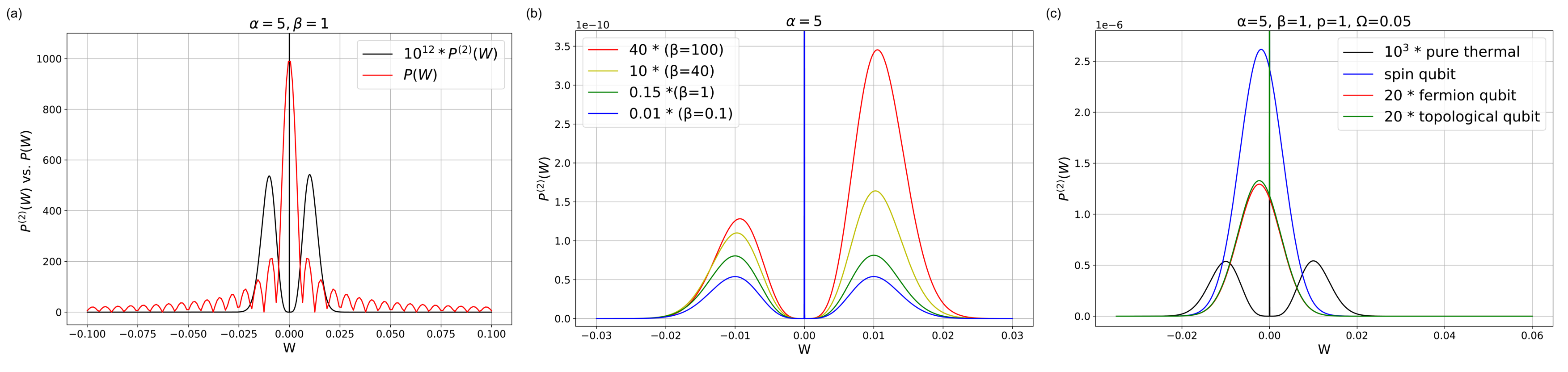}}
    \caption{Example WDFs: (a) comparison of $P(W)$ and $P^{(2)}(W)$ for pure thermal bath; (b) $P^{(2)}(W)$ for pure thermal bath by varying $\beta$; (c) $P^{(2)}(W)$ with qubit coupling to thermal bath. The sticks at $W=0$ represent $p_0 \delta(W)$ of $P^{(2)}(W)$.
}
\label{prl_pw}
\end{figure*}

The above results are up to ${\cal O}(\lambda^2)$. However, we can obtain the nonperturbative WCF for the thermal states. This can be achieved with the help of the following generalized Wick's theorem \cite{vanLeeuwen:2011gi, Diosi:2017kzh, Ferialdi:2021ywo, Ferialdi:2023bjb} relating the orderings $\mathbb{O}$ and $\mathbb{O}'$:
\be\label{Wick_thm_m}
\mathbb{O}\Big[ e^{\int  \lambda(x) V(x) d^4x } \Big] =e^{{1\over 2}\int\int  \lambda(x) \lambda(y) C(x,y) d^4x d^4y} \; \mathbb{O}'\Big[ e^{\int  \lambda(x) V(x) d^4x} \Big]
\ee
where $C(x,y)=(\mathbb{O} - \mathbb{O}') V(x) V(y)$
is a pure commutator and thus a c-number, as long as the 
ordering bases of $\mathbb{O}$ and $\mathbb{O}'$ are linearly related. For our purpose, we will choose $\mathbb{O}=\{\cal{I}, \cal{T}_{\pm}\}$ and $\mathbb{O}'$ to be the normal ordering $\cal{N}_{\beta}$ for the thermal state $\rho_{\beta}$ \cite{Evans:1996bha, Blasone:1997nt, Evans:1998yi}. We can verify $C(x,y)={\rm Tr}\big[\rho_{\beta}\mathbb{O}V(x) V(y)\big]$. Applying the general Wick's theorem \eq{Wick_thm_m} associated with $\cal{N}_{\beta}$ twice to simplify \eq{chi_realtime_m}, we obtain nonperturbative WCF for thermal state $\rho_{\beta}$ to all orders of $\lambda$ \footnote{Although the application of generalized Wick's theorem to derive the nonperturbative WCF is quite fascinating, it is technical and less relevant to the main theme of this paper. In Appendix \ref{app_wick}, we have given a thorough review of the generalized Wick's theorem and detailed the steps of deriving \eq{chiv_all_m}.},
\be\label{chiv_all_m}
\chi(v)=e^{\chi^{(2)}(v)-1} {\rm Tr}\Big[\rho_{\beta} {\cal N}_{\beta} e^{\int_{-\infty}^{\infty} dt \big(\lambda^+(t;v) -\lambda^-(t;v) \big) V_I(t)} \Big] = e^{\chi^{(2)}(v)-1} 
\ee
where $\chi^{(2)}(v)$ is given by \eq{chiv_omega_1_m} and \eq{KMS_m}. The second identity is achieved by ${\cal N}_{\beta}$'s defining property. From \eq{chiv_all_m} and $\chi^{(2)}(i\beta)=1$, we find that the integrated form of JC-FT holds to all orders of $\lambda$, i.e., $\chi(i\beta)=1$, as expected. By \eq{W_ext}, we find that $W_{\rm ext} = W^{(2)}_{\rm ext} \; e^{\chi^{(2)}(0)-1} = W^{(2)}_{\rm ext}$. However, the higher moments of $W$ will deviate from the ${\cal O}(\lambda^2)$ ones.

Besides, we can Fourier transform $\chi(v)$ to obtain $P(W)$. There is no closed form for $P(W)$. However, one can obtain it either by numerical Fourier transform or by the order-by-order form for any order of $\lambda$. It is interesting to compare this non-perturbative $P(W)$ with $P^{(2)}(W)$ of \eq{pW_1_m} for some chosen spectral functions along with the driven source of \eq{lambdaW}. Using the spectral function of \eq{S_ohmic}, we show such comparison for $\alpha=5, \beta=1$ in Fig. \ref{prl_pw}(a). We see that the full $P(W)$ exhibits small oscillations, which are absent in $P^{(2)}(W)$. As far as we know, the full $\chi(v)$ and $P(W)$ for the thermal states have not been obtained in the literature.

\section { Work statistics of thermal bath with qubit coupling}\label{sec4}
We will now consider the work statistics for the thermal bath coupled to a spin or fermion qubit, i.e., the driven protocol \eq{H_tot_m} with $H_0$ and $V$ given by
\be
H_0 = H_0^{\rm TB} + H_0^{\rm Q} \qquad {\rm with} \quad H_0^{\rm Q} =\Omega |1\rangle\langle 1|
\ee
and 
\begin{align}\label{qubit_V}
V = \begin{cases}
    \sigma_x \otimes O \;,  & \mbox{spin coupling}\;, \\
    d^{\dagger}\Psi + \Psi^{\dagger} d = -i \sum_{m=1,2} \gamma_m O_m\;,  & \mbox{fermion coupling}   
  \end{cases} \qquad
\end{align}
acting on the initial diagonal state
\be
\rho = \rho_{\rm Q} \otimes \rho_{\beta} \quad \mbox{with} \quad \rho_{\rm Q}=p |0\rangle\langle 0| + (1-p) |1\rangle\langle 1|
\ee
with $p\in [0,1]$. Here, $|0\rangle$ and $|1\rangle$ are, respectively, the ground and excited states of the qubit. The qubit state $\rho_{\rm Q}$ can be thought of as a Boltzmann thermal state with the inverse temperature given by 
\be\label{beta_Q}
\beta_{\rm Q} := -{1\over \Omega} \ln {1-p \over p}\;.
\ee
Naively, the qubit in a thermal state can be achieved by putting it in contact with a thermal bath at the specified (positive) temperature. In reality, controlling the qubit's temperature when verifying the fluctuation theorem is a key challenge; see recent experiments on a quantum computer in \cite{solfanelli2021experimental, hahn2023quantum, BassmanOftelie:2024rsn}. Moreover, $\beta_{\rm Q} <0$ if $p<1/2$. In such cases, the qubit state exhibits population inversion, which should be maintained by external pumping, for example, by a laser. Thus, this qubit state is not passive, and the consideration of work extraction is incomplete without taking into account the required work to achieve such negative-temperature states. Below, we will restrict ourselves to the regime with positive $T_Q=1/\beta_Q$ when considering the qubit-bath system as a heat engine/refrigerator. Otherwise, the corresponding efficiency or COP will be higher than the Carnot ones due to the neglect of the external pumping power.

For the spin coupling, $\sigma_x:=|0\rangle\langle 1|+ |1\rangle\langle 0|$ and $O$ is a bosonic field operator of thermal bath. For the fermion coupling, $d=|0\rangle\langle 1|$, and $\Psi$ is a fermion field operator of the thermal bath. Moreover, we decompose a fermion into a pair of Majorana zero modes (MZMs) as follows:
\be
d={\gamma_1 + i \gamma_2 \over 2}\;, \quad \Psi= O_2 - i O_1\;.
\ee
Note that MZMs are hermitian and satisfy the Clifford algebra, e.g., $\gamma_m^{\dagger}=\gamma_m$ and $\{\gamma_m, \gamma_n\}=2 \delta_{m,n}$. If we consider the fermion qubit, this decomposition is just formal for the convenience of calculation. However, we can also consider the topological qubit \cite{ho2014decoherence, liu2016decoherence} with $\gamma_1$ and $\gamma_2$ being far separated so that the thermal correlators of the far-separating hermitian operators $O_1$ and $O_2$ obey cluster decomposition by causality constraint \cite{Brown_1992}. i.e.,
\be\label{cluster_p_m}
\langle (O_1)^k  (O_2)^l \rangle_{\beta} \simeq \langle (O_1)^k \rangle_{\beta} \langle (O_2)^l \rangle_{\beta} \delta_{k, 2 {\bf N}} \delta_{l, 2 {\bf N}}\;.
\ee
Since the qubit coupling is quadratic for basis change, the relation \eq{Wick_thm_m} does not hold. Thus, there is no simple way to evaluate the full $\chi(v)$, and we will only restrict to ${\cal O}(\lambda^2)$.

To evaluate $\chi^{(2)}(v)$ and $P^{(2)}(W)$ given by \eq{chiv_omega_1_m} and \eq{p_W_1_m}, we need to calculate the real-time Green functions of $V_I(t):=e^{i H_0 t} V e^{-i H_0 t}$ in the state $\rho$. Since $\rho$ is a product state, the real-time Green functions also take the product form. Thus, their Fourier transforms take the convolution forms of the qubit's and thermal ones. The calculations are tedious but straightforward. We omit the details here, but the readers can find them in Appendix \ref{app_C} for the spin qubit coupling and Appendix \ref{app_D} for the fermion and topological qubit coupling. We summarized the results in a unified form as follows: 
\begin{align} 
& i \tilde{G}^{+-}(\omega) = e^{-\beta \omega} \Big[ p e^{-\beta \Omega} S^{(1)}_+(\omega) + (1-p)  e^{\beta \Omega}  S^{(2)}_-(\omega )  \Big]\;, \nn
\\
& i \tilde{G}^{-+}(\omega) = p S^{(1)}_-(\omega)+ (1-p) S^{(2)}_+(\omega)\;.  
\label{UDW_G_Q}
\end{align}
In the above, we have introduced the short-handed notations: $S^{(i=1,2)}_{\mp}(\omega):= S^{(i=1,2)}_{\beta}(\omega \mp \Omega)$.
Using \eq{UDW_G_Q}, \eq{chiv_omega_1_m} and \eq{W_ext_f}, we can obtain 
\bea
\chi^{(2)}(i\beta) &=& 1 - {1\over 2}\int_{\lambda} (1-e^{-\beta \omega}) \; \Big[p \big(S^{(1)}_-(\omega)- e^{-\beta \Omega} S^{(1)}_+(\omega) \big)  \nn \\ 
 && \qquad + \; (1-p) \big( S^{(2)}_+(\omega)- e^{\beta \Omega} S^{(2)}_-(\omega)\big) \Big]\;, \nn
\\
\overline{W}^{(2)} &=& - {1\over 2} \int_{\lambda} \omega \; \Big[ p \big( S^{(1)}_-(\omega)- e^{-\beta(\omega + \Omega)} S^{(1)}_+(\omega)\big) 
\nn \\ \label{chi2_W2_exp}
&& \qquad + \;  (1-p) \big( S^{(2)}_+(\omega)- e^{-\beta(\omega -\Omega)} S^{(2)}_-(\omega)\big) \Big]\;. \qquad  
\eea

The spectral functions $S^{(i=1,2)}_{\beta}(\omega)$ in the above are specified as below for different qubit couplings:
\begin{enumerate}

 \item for spin coupling:  $S^{(1)}_{\beta}(\omega)=S^{(2)}_{\beta}(\omega)$ are the Wightman spectral function of neutral scalar quasiparticle $O$ in state $\rho_{\beta}$, i.e., they equal to $e^{\beta \omega} n_{\rm BE}(\omega) S^O(\omega)$;
 
\item for fermion coupling: $S^{(1)}_{\beta}(\omega)=n_{\rm FD}(\omega)S^{\Psi}(\omega)$  and $S^{(2)}_{\beta}(\omega)= e^{\beta \omega}  n_{\rm FD}(\omega) S^{\Psi}(\omega)$ are respectively the spectral densities of quasiparticles and quasiholes, where $n_{\rm FD}(\omega)=(e^{\beta \omega}+1)^{-1}$ and $S^{\Psi}(\omega) :=\int_{-\infty}^{\infty} dt e^{i\omega t} \{\Psi^{\dagger}(t), \Psi(0) \}$ is the Paui-Jordan spectral function of $\Psi$. To ensure thermal states' stability, $S^{\Psi}(\omega)$ should be bounded below; a

\item for topological qubit coupling with the causality constraint \eq{cluster_p_m}: both $S^{(1)}_{\beta}(\omega)=S^{(2)}_{\beta}(\omega)= {1\over 2} S^{\Psi}(\omega)$, which are  $\beta$-independent. 
\end{enumerate}
Note that \eq{UDW_G_Q} is reduced to \eq{KMS_m} for gapless qubit, i.e., $\Omega=0$ if $S^{(1)}_{\beta}(\omega)=S^{(2)}_{\beta}(\omega)$, so that they behave as the passive thermal state.  Otherwise, the thermality or passivity can be violated.

Substituting the real-time Green functions in \eq{p_W_1_m} and \eq{m_p0} by \eq{UDW_G_Q}, we can obtain $P^{(2)}(W)$ for the three cases of qubit coupling to the thermal bath. 
In Fig. \ref{prl_pw}(b) and (c), we compare the $P^{(2)}(W)$ for the thermal states without and with qubit coupling for the driven source \eq{drive_s} and the Ohmic-type spectral function of \eq{S_ohmic} for $V=O, \Psi$. In Fig. \ref{prl_pw}(b), the WDFs tend to be asymmetric about $W=0$ for larger $\beta$. These WDFs yield $W^{(2)}_{\rm ext} =- \overline{W}^{(2)} \le 0$, consistent with the passivity of pure thermal baths. On the other hand, in Fig. \ref{prl_pw}(c), we show an example of qubit coupling to the thermal bath, which yields $W_{\rm ext}>0$. Moreover, we can also define the energy gain by qubit coupling as $E_{\rm gain}:=W_{\rm ext} -{\rm Tr}(\rho_{\rm Q} H_0^{\rm Q})=W_{\rm ext}-(1-p)\Omega$. Since $p=1$ in the example of Fig. \ref{prl_pw}(c), it yields $E_{\rm gain}>0$ though its size is rather small due to small $P^{(2)}(W\ne 0)$. The WDF generally depends on the parameters $\alpha$, $\beta$, $p$, and $\Omega$. More examples of WDF, such as the ones with sub-Ohmic ($\alpha<1$) and Ohmic ($\alpha=1$) spectral functions and with either $p\approx 1$ or $p\approx 0$, are given in Appendix \ref{app_E1}. Two more interesting points observed from these plots: (i) $P^{(2)}(W)$'s of the pure thermal bath are bimodal if $\alpha>1$ but unimodal otherwise. However, the bimodal $P^{(2)}(W)$ will turn unimodal when changing to qubit coupling, as shown in Fig. \ref{prl_pw}(c).  (ii) The overall magnitude of $P^{(2)}(W)$ for the spin qubit coupling is higher than that of fermion or topological qubit couplings. This is due to the difference between the thermal factors for the bosonic and fermionic field operators, to which the spin and fermion qubits couple, respectively. The bosonic thermal factor favors a larger low-energy population than the fermionic one for the same spectral density. With a given ultraviolet cutoff on the spectral density, the low-energy modes dominate in facilitating the energy transfer, and cause the larger magnitude of WDF for the spin qubit coupling as shown in Fig. \ref{prl_pw}(c). This is the statistical effect of the identical particles on the work statistics.

\section { Work extraction by qubit coupling and its efficiency }\label{sec5}

\begin{figure*}[hbt!]
    \centerline{\includegraphics[width=1.1\linewidth]{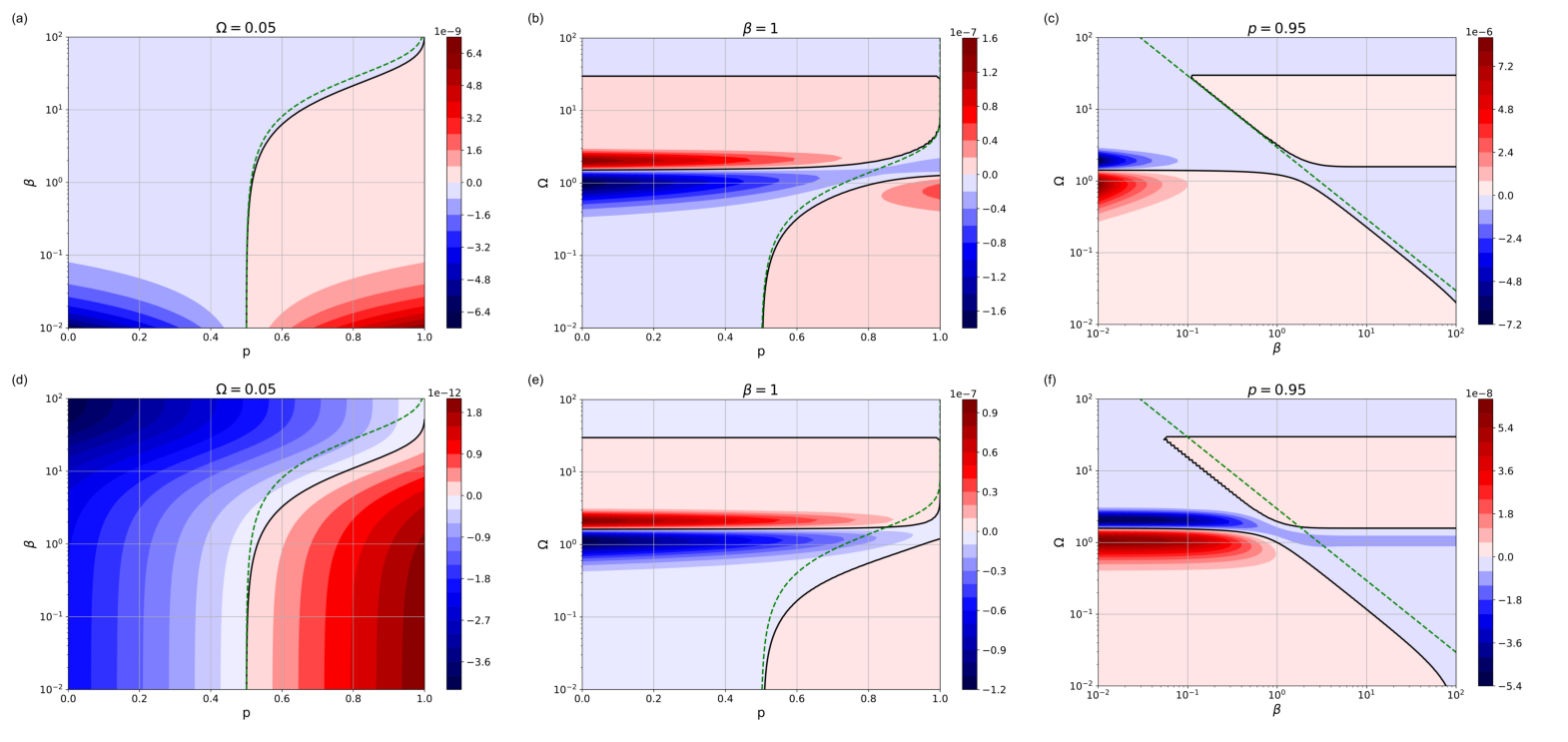}}
    \caption{Density plots of $W^{(2)}_{\rm ext}$ for spin qubit coupling in subfigures (a)-(c), and fermion qubit coupling in subfigures (d)-(f). All are for the super-Ohmic spectral function with $\alpha=5$ but with one of three parameters $\Omega$, $\beta$, or $p$ fixed. The plots' $\beta$ and $\Omega$ are presented on a log scale for clarity.   $W^{(2)}_{\rm ext}=0$ represented by the black lines  serve as the boundary between $W^{(2)}_{\rm ext}>0$ (red gradient) and $W^{(2)}_{\rm ext}<0$ (blue gradient) regimes. The green dashed lines denote  $\beta=\beta_{\rm Q}$ with $\beta_Q$ given by \eq{beta_Q}. 
}
\label{prl_w_ext}
\end{figure*}

\begin{figure*}[hbt!]
   \centerline{\includegraphics[width=1.1\linewidth]{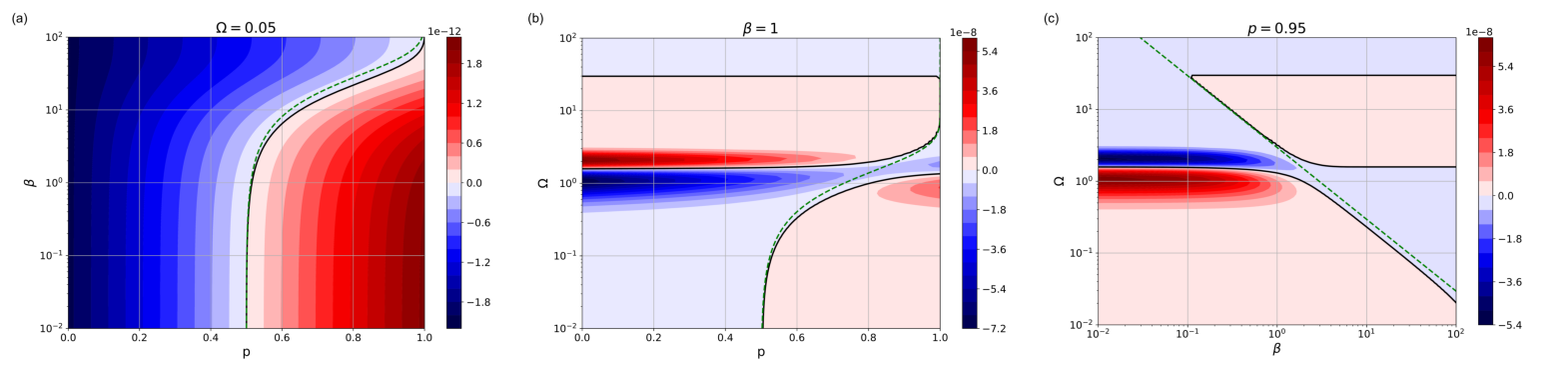}}
    \caption{Similar density plots of $W^{(2)}_{\rm ext}$ to the ones in Fig. \ref{prl_w_ext} but for the topological qubit coupling. Unlike the case of fermion qubit coupling, the $W^{(2)}_{\rm ext} = 0$ line does not intersect with the $\beta = \beta_Q$ line.
    }
    \label{prl_w_ext_sup_to}
    \end{figure*}

The WDFs in Fig. \ref{prl_pw}(c) suggest that there exists some physical regime with $W^{(2)}_{\rm ext}>0$ by the qubit coupling. In this case, the qubit-bath system acts as a heat engine.  On the other hand, it acts as a refrigerator for the regime with $W^{(2)}_{\rm ext} < 0$.  Even though the quantum heat engine is intriguing, the quantum refrigerator is also essential for suppressing decoherence and reducing gate errors in noisy intermediate-scale quantum (NISQ) devices.
This motivates the full survey of the (sign) distribution of $W^{(2)}_{\rm ext}$ in the parameter space of $\beta$, $p$, and $\Omega$. The $W^{(2)}_{\rm ext}$ for qubit coupling can be obtained  via \eq{W_ext_f} by evaluating $S^{V=O,\Psi}(\omega)$ therein with \eq{UDW_G_Q} and \eq{S_ohmic} for the  driven source \eq{drive_s}.  In Fig. \ref{prl_w_ext}, we give the 2D density plots of $W^{(2)}_{\rm ext}$ for the spin qubit coupling in the subfigures (a)-(c) and the fermion qubit coupling in the subfigures (d)-(f).

From Fig. \ref{prl_w_ext}, we first note that the overall size of $W^{(2)}_{\rm ext}$ for the spin qubit case is about one or two orders higher than the one for the fermion qubit case, which is also implied by the corresponding overall sizes of $P^{(2)}(W)$ shown in Fig. \ref{prl_pw}(c). This could be due to more coherent expectations for bosonic quasiparticles than for the fermionic ones. It also implies that work extraction, if possible, is more efficient through spin qubit coupling than the fermion one. Secondly, the work extraction is expected as the qubit state can act as a small "thermal bath" with inverse temperature $\beta_{\rm Q}$ of \eq{beta_Q}. Moreover, due to the mismatch of thermal statistics, i.e., Boltzmann weight for the qubit and Bose-Einstein or Fermi-Dirac weights for the thermal bath, the qubit still cannot equilibrate with the thermal bath even $\beta_{\rm Q} =\beta$. This can explain why the black lines for $W^{(2)}_{\rm ext}=0$ in Fig. \ref{prl_w_ext} do not coincide with the green dashed lines for $\beta=\beta_{\rm Q}$. Thirdly, despite the similarity in the overall features of the plots for the spin and fermion qubit couplings, the black lines and green dashed lines never cross each other for the former case, but they do for the latter. This shows the difference between the bosonic and fermionic quasiparticles from work statistics. Finally, we find that $W^{(2)}_{\rm ext} > 0$ happens at least for the regime with small $\Omega$, larger $p$, and smaller $\beta$, denoted as the regular regime of work extraction. But the subfigures (b), (c), (e), and (f) of Fig. \ref{prl_w_ext} also indicate some exceptions. In particular, the size of $W^{(2)}_{\rm ext}$ for these exception regimes is several orders higher than that of the regular regime.

\begin{figure*}[hbt!]
	\centering
    \includegraphics[width=0.9\linewidth]{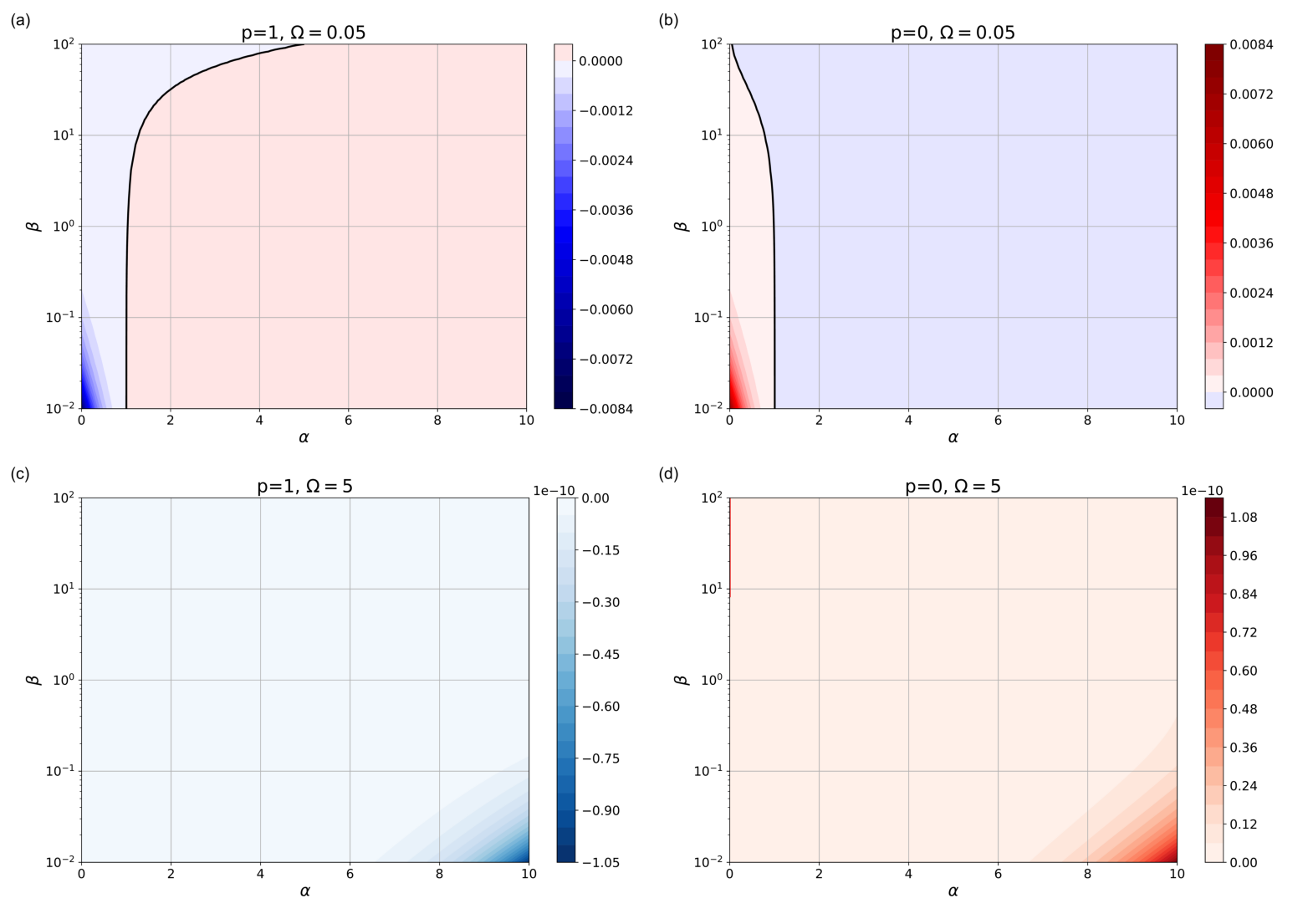}
    \caption{Density plots of $W^{(2)}_{\rm ext}$ on the $(\alpha,\beta)$-plane for spin qubit coupling for (a) $p=1, \Omega=0.05$, (b) $p=0, \Omega=0.05$, (c) $p=1, \Omega=5$ and (d) $p=0, \Omega=5$. The black lines  represent $W^{(2)}_{\rm ext} = 0$, separating regions where $W^{(2)}_{\rm ext} > 0$ (red gradient) and $W^{(2)}_{\rm ext} < 0$ (blue gradient). Note that $\beta$ is presented on a log scale.
    }
    \label{prl_w_ext_sup_ab}
\end{figure*}

\begin{figure*}[hbt!]
	\centering
    \includegraphics[width=0.9\linewidth]{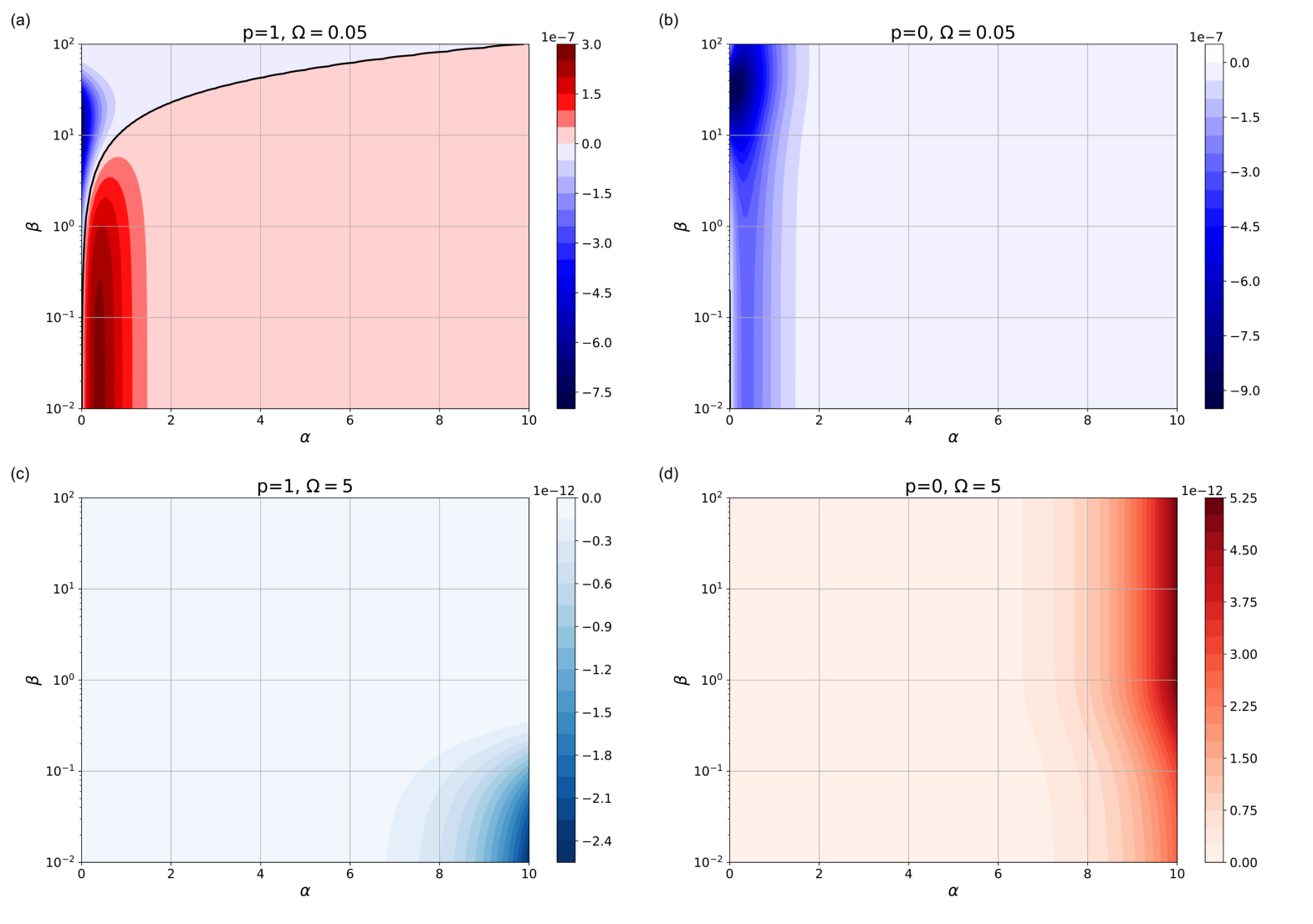}
    \caption{Similar plots to Fig. \ref{prl_w_ext_sup_ab} but for fermion qubit coupling. In the subfigure (b), the $W^{(2)}_{\rm ext}>0$ only occurs at the margin of $\alpha=0$.
    }
    \label{prl_w_ext_sup_af}
\end{figure*}

The $W^{(2)}_{\rm ext}$ for the topological qubit coupling bears similar features of fermion qubit coupling, as shown in Fig. \ref{prl_w_ext_sup_to}. However, there is a subtle difference: unlike the case of fermion qubit coupling, here the $W^{(2)}_{\rm ext} = 0$ line does not intersect with the $\beta = \beta_Q$ line.

In the above, we have considered the $W^{(2)}_{\rm ext}$ plots with fixed $\alpha=5$. We have presented the density plots of $W^{(2)}_{\rm ext}$ on the planes of $(p,\beta)$, $(p,\Omega)$ and $(\beta,\Omega)$ for the spin and fermion qubit couplings in Fig. \ref{prl_w_ext}. To complete the full survey, we also present the density plots of $W^{(2)}_{\rm ext}$ on $(\alpha,\beta)$-plane in Fig. \ref{prl_w_ext_sup_ab} (Fig. \ref{prl_w_ext_sup_af}) for the spin (fermion) qubit coupling for (a) $p=1, \Omega=0.05$, (b) $p=0, \Omega=0.05$, (c) $p=1, \Omega=5$ and (d) $p=0, \Omega=5$.

These plots imply the following: (i) $W^{(2)}_{\rm ext}>0$
($W^{(2)}_{\rm ext}<0$) regime spans nearly entire $(\alpha,\beta)$-plane for $p=0$ ($p=1$) if $\Omega$ is large enough, but the regime becomes restricted as $\Omega$ decreases; (ii) The $W^{(2)}_{\rm ext}>0$ regime for $p=0$ ($W^{(2)}_{\rm ext}<0$ for $p=1$) is restricted to the regime small $\alpha$; (iii) as $p\simeq 0$ is changed to $p\simeq 1$, the corresponding $W^{(2)}_{\rm ext}>0$ and $W^{(2)}_{\rm ext}<0$ regimes roughly swap. This is consistent with (i) and (ii). 

 Supplementary to the above density plots of $W^{(2)}_{\rm ext}$, we also present in Appendix \ref{app_E2} more pedagogical plots about the dependence of $W^{(2)}_{\rm ext}$ on either $\beta$ or $p$ for all three qubit-coupling cases. These plots give more direct comparisons among three qubit-coupling cases on the patterns of $W^{(2)}_{\rm ext}$.

\begin{figure*}[htb!]
    \centerline{\includegraphics[width=1.1\linewidth]{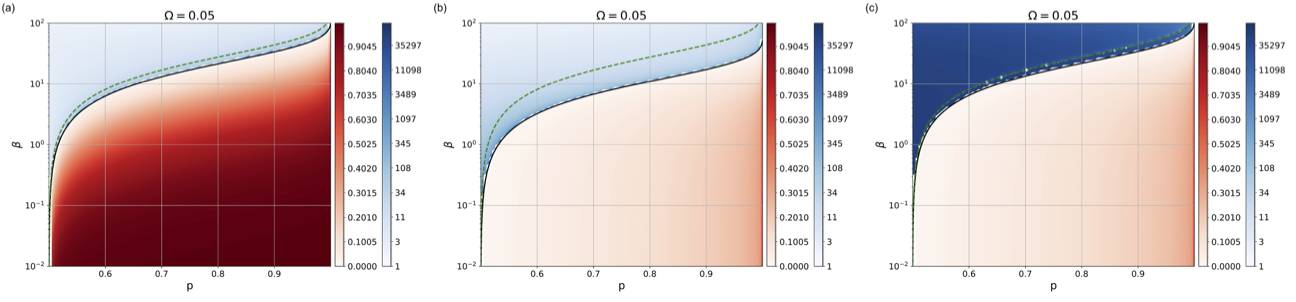}}
    \caption{Density plots of the efficiency (red-like color) and COP (blue-like color) of the qubit-bath system shown in Fig. \ref{fig_scheme}(b) operating as a heat engine ($\overline{W}^{(2)}<0$) or a refrigerator ($\overline{W}^{(2)}>0$). The subfigure (a) is for the spin qubit, (b) for the fermionic qubit, and (c) for the topological qubit.  
    The results of (a) and (b) are the counterparts of Fig. \ref{prl_w_ext} and Fig. \ref{prl_w_ext_sup_to} for work extraction. In each subfigure, the first color bar is for the efficiency, and the second one for the COP; the black line corresponds to $\overline{W}^{(2)} = 0$, and the green-dotted line to $\beta=\beta_Q$. The parameter range of $p$ is $[0.5 + \epsilon, 1 - \epsilon]$ with $\epsilon = 10^{-6}$, which nearly corresponds to $\beta_Q \in [0, \infty]$.}
\label{prl_eff_cop}
\end{figure*}

Finally, we consider the setup in Fig. \ref{fig_scheme}(b) as the heat engine/refrigerator, and its efficiency $\eta$ or COP are give by \eq{efficiency} and \eq{COP}, respectively, in terms of $\overline{W}^{(2)}$ (or $W_{\rm ext}^{(2)}=- \overline{W}^{(2)}$),  $\Delta S=\beta \overline{W}^{(2)} + \ln \chi^{(2)}(i\beta)$, $T_B=1/\beta$ and $T_Q=1/\beta_Q$ with $\beta_Q$ defined in \eq{beta_Q}. The explicit forms of $\overline{W}^{(2)}$ and $\chi^{(2)}(i\beta)$ are given by \eq{chi2_W2_exp}.  Moreover, we will only consider the regime of $T_Q>0$ (or $p>0.5$) for the reason discussed earlier.  The qubit-bath system of Fig. \ref{fig_scheme}(b) behaves as a heat engine if $\overline{W}^{(2)}<0$, i.e., $W_{\rm ext}>0$, and as a refrigerator if $\overline{W}^{(2)}>0$. In Fig. \ref{prl_eff_cop}, we show the density plots of the efficiency $\eta$ for the regime with $\overline{W}^{(2)}<0$ or COP for $\overline{W}^{(2)}>0$. The subfigures of Fig. \ref{prl_eff_cop} are the counterparts of Fig. \ref{prl_w_ext} and Fig. \ref{prl_w_ext_sup_to} for $W^{(2)}_{\rm ext}$ of work extraction for coupling of spin/fermionic/topological qubit to the bath under cyclic driving processes.

The results shown in Fig. \ref{prl_eff_cop} convey some interesting information:

(i) As observed in Figs. \ref{prl_w_ext},  \ref{prl_w_ext_sup_to}, \ref{prl_eff_cop}, the black lines for $\overline{W}^{(2)}=0$  generally do not coincide with the green dotted lines for $\beta=\beta_Q$. This implies that even if the qubit and the bath have the same temperature, i.e., they could reach thermal equilibrium in the classical thermodynamic sense, the average work done $\overline{W}^{(2)}$ can still be nonzero, so the qubit-bath system can still behave as a heat engine or a refrigerator. 

(ii) The results of density plots are consistent with our expectation for the efficiency or COP as in classical thermodynamics: a larger $T_H/T_L$ yields a larger $\eta$ but a smaller COP. As the ratio approaches unity, $\eta \rightarrow 0$ and COP$\rightarrow \infty$. The latter result is indicated by the dashed roughness of the blue color on top of the black line. 

(iii) Comparing the color densities of the three subfigures in Fig. \ref{prl_eff_cop}, we can see how the quantum statistics affect the efficiency of a heat engine or the COP of a refrigerator. This is similar to how the amount of the work extraction depends on the quantum statistics of the qubit and the bath quasiparticle operators. We see that the bosonic statistics of the spin-qubit generally yield a far larger efficiency of the heat engine than the other two alternatives. On the other hand, the topological qubit generally yields a far larger COP of the refrigerator than the other two.  This can serve as a guideline for designing a quantum heat engine and refrigerator, taking into account the underlying quantum statistics.

\section { Conclusion}\label{sec6}

In this paper, we have adopted the effective field theory approach to calculate the work statistics of an irreversible process without and with qubit coupling to the thermal bath. With such an efficient tool, we can explicitly compute the work characteristic function, work distribution function, and possible work extraction by varying model parameters, such as the type of Ohmic spectral function for the quasiparticles, the temperature of the thermal bath, and the spectral gap of the qubit system. In particular, with our method, we can derive the nonperturbative work distribution for thermal states. As far as we know, this is the first nonperturbative result obtained in the literature. We also find that the work extractable regime via qubit coupling is usually restricted, but can be precisely pinned down from our framework's results. This will be useful for designing the quantum engine. Furthermore, we can treat the qubit-bath system as a heat engine or refrigerator and derive the corresponding efficiency or COP from work statistics by assuming first-law relations for adiabatic processes. Our results indicate that the spin qubit coupling yields a far more efficient heat engine than the other two alternatives, but the topological qubit coupling yields a far better refrigerator. This implies that the underlying quantum statistics of the qubit-bath system are highly relevant to the design of a quantum heat engine or refrigerator. 

Our method and results extend the fluctuation theorem and quantum thermodynamics to many-body systems beyond pure thermal states. Further extension of our method by including the battery system can improve the study of quantum heat engines as a practical application of the fluctuation theorem and give explicit examples of some statements implied by the resource theory.

\bigskip

\noindent {\it Acknowledgements.}
This work is supported by Taiwan's National Science and Technology Council (NSTC) through Grant No.~112-2112-M-003-006-MY3. We thank the anonymous referee of PRE for the inspiring suggestion of considering the efficiency of a quantum engine using the first law relations.

\appendix

\bigskip

\section{Review of work statistics in real-time formalism}\label{app_A}

In this Appendix, we review the work statistics in real-time formalism as first considered in \cite{talkner2007fluctuation1, fei2020nonequilibrium} by illustrating the subtle details. 
We consider the work statistics done by an external agent during an irreversible process. In particular, we consider the irreversible process generated by a Hamiltonian with a time-dependent coupling constant $\lambda(t)$:
\be\label{H_tot}
H(t) = H_0 + \lambda(t) V
\ee
where $H_0$ and $V$ are time-independent and hermitian operators, and $H_0$ is usually the term of kinetic energy, and $\lambda(t)$ is the time-dependent source function to excite operator $V$. Since the total Hamiltonian is time-dependent, the evolution is best described in the Schwinger-Keldysh formalism with the results expressed in terms of the real-time Green functions. We will then use this formalism to study the work statistics for the irreversible process dictated by the evolution with the Hamiltonian \eq{H_tot}. We will denote the initial state of the system by $\rho$.

The work $W$ performed by the external agent is a random variable. Thus, we can introduce work statistics characterized by a probability distribution $P(W)$ with $\int_{-\infty}^{\infty} dW P(W)=1$. Although $W$ is not physically sensible, is its average 
\be
\overline{W}:=\int_{-\infty}^{\infty} dW W P(W)
\ee
and its cousin cumulants. In the context of the fluctuation theorem \cite{crooks1999entropy, crooks2000path, jarzynski1997nonequilibrium, seifert2012stochastic}, one can introduce the generating function for the cumulants of the work statistics, called the work characteristic function (WCF), and given by
\be
\chi(v)= \int_{-\infty}^{\infty} dW e^{i v W} P(W)\;. 
\ee
By inverse transform, we get 
\be\label{PW_def}
P(W)=\int_{-\infty}^{\infty} {dv \over 2\pi} e^{-i v W} \chi(v)\;.
\ee
Note that
\be
\chi(0)=1
\ee
and 
\be\label{W_ave_def}
\overline{W} := \int_{-\infty}^{\infty}  W P(W) dW = -i \lim_{v\rightarrow 0} {\partial \chi(v) \over \partial v}\;.
\ee
and the higher cumulants in a similar way.

In \cite{talkner2007fluctuation1}, the WCF for an irreversible process starting from time $t_i$ and ending at time $t_f$ can be written as 
\be\label{chi_def}
\chi(v)={\rm Tr}\big[ U_S(t_f,t_i) e^{-i H(t_i) v} \; \rho \; U_S(t_i, t_f) e^{iH(t_f) v} \big]
\ee
where the time-ordered evolution operator in the Schrödinger picture is denoted by $U_S(t_2,t_1)$, which represents 
${\cal T}_+ \big[ e^{-i \int_{t_1}^{t_2} H(t) dt}\big]$ if $t_2\ge t_1$ or ${\cal T}_-\big[ e^{-i \int_{t_1}^{t_2} H(t) dt} \big]={\cal T}_-\big[ e^{i \int_{t_2}^{t_1} H(t) dt} \big]$ if $t_1 \ge t_2$. Here, ${\cal T}_+$ and ${\cal T}_-$ are the time ordering symbols for the forward and backward processes, respectively. Moreover, the derivation of \eq{chi_def} requires \cite{talkner2007fluctuation1}
\be\label{cond_rho}
[H(t_i), \rho]=0\;.
\ee
If we consider the cyclic process with $H(t_f)=H(t_i)$, then \eq{chi_def} takes the form of an out-of-time order correlator (OTOC), i.e., $\langle W_1^{\dagger}(t_f-t_i) W_2^{\dagger} W_1(t_f-t_i) W_2 \rangle$ with $W_1(t_f-t_i) = U_S(t_i,t_f)$ and $W_2=e^{i H(t_i) v}$. This was also noted in \cite{yunger2017jarzynski}.

We want to express the WCF $\chi(v)$ of the work statistics in terms of real-time Green functions. To proceed, we first extend the usual real-time contour on the time interval $[t_i, t_f]$ into the extended one $[t_i, t_f + v]$ as shown in Fig. \ref{contour} in the main text. Secondly, we need to transform from the Schrödinger picture to the interaction picture by
\be\label{U_I}
U_S(t_2,t_1)=e^{-i H_0 t_2}   {\cal T}_{\pm} \big[ e^{-i \int_{t_1}^{t_2} \lambda(t) V_I(t) dt} \big]  e^{i H_0 t_1}
\ee
with the choice of ${\cal T}_+$ for $t_2\ge t_1$ or ${\cal T}_-$, otherwise. The operator $V_I(t)$ is the counterpart of $V$ in the interaction picture given by 
\be
V_I(t)=e^{i H_0 t} V e^{-i H_0 t}\;.
\ee
Using this, we have
\begin{align}
    U_S(t_i, t_f) =& \,e^{- i H_0 t_i}{\cal T}_- \big[ e^{- i \int_{t_f}^{t_i} \lambda(t) V_I(t) dt} \big]  e^{i H_0 t_f},\\ 
U_S(t_f, t_i) =& \,e^{- i H_0 t_f} {\cal T}_+ \big[ e^{-i \int_{t_i}^{t_f} \lambda(t) V_I(t) dt} \big]  e^{i H_0 t_i}\nonumber\\ 
=& \,e^{- i H_0 t_f} {\cal T}_+ \big[ e^{-i \int_{t_i+v}^{t_f +v} \lambda(t-v) V_I(t-v) dt} \big] e^{i H_0 t_i} \;. \label{U_I_int}
\\
=&  \,e^{- i H_0 (t_f +v)} {\cal T}_+ \big[ e^{-i \int_{t_i + v}^{t_f +v} \lambda(t-v) V_I(t) dt} \big] e^{i H_0 (t_i + v)}\;. \label{U_I_c}
\end{align}
When transforming $U_S(t_f,t_i)$, we first shift the time interval $[t_i,t_f]$ to $[t_i+v, t_f +v]$ in \eq{U_I_int}, and then apply
\begin{align}
   & V_I(t_1-v) V_I(t_2-v
) \cdots V_I(t_n-v)\nonumber\\
=&\, e^{-i H_0 v} \big[ V_I(t_1) V_I(t_2) \cdots V_I(t_n) \big] e^{i H_0 v} 
\end{align}

to arrive \eq{U_I_c}.

Similarly, on the extended part of the contour in Fig. \ref{contour}, by using \eq{U_I} again, we have
\begin{align}
    e^{i H(t_f) v} &= e^{-i H_0 t_f } {\cal T}_- \big[ e^{-i \lambda_f \int_{t_f +v}^{t_f}   V_I(t) dt} \big]  e^{i H_0 (t_f +v)} \nonumber\\
    &= e^{-i H_0 t_f } {\cal T}_- \big[ e^{i \lambda_f \int^{t_f +v}_{t_f}   V_I(t) dt} \big]  e^{i H_0 (t_f +v)}\;,\\
    e^{- i H(t_i) v} &=  e^{-i H_0 (t_i + v)} {\cal T}_+ \big[ e^{-i \lambda_i \int_{t_i}^{t_i+v}   V_I(t) dt} \big] e^{i H_0 t_i} \;.  
\end{align}
with
\be
\lambda_i:=\lambda(t_i)\;, \qquad \lambda_f:=\lambda(t_f)\;.
\ee

Apply the above transformations from the Schrödinger picture to the interaction picture to \eq{chi_def}; we can obtain
\begin{align}\label{chiv_11}
    \chi(v) =& \, {\rm Tr} \Big\{ {\cal T}_+\big[e^{ -i \int_{t_i + v}^{t_f +v} \lambda(t-v) V_I(t) dt }  \big] {\cal T}_+\big[e^{-i \lambda_i \int_{t_i}^{t_i + v}   V_I(t) dt}  \big]  \; \rho_I \; \times \nonumber\\
    &{\cal T}_-\big[e^{i \int^{t_f}_{t_i} \lambda(t) V_I(t) dt }  \big]{\cal T}_-\big[e^{ i \lambda_f \int^{t_f +v}_{t_f}   V_I(t) dt}  \big] \Big\} \;. 
\end{align}
with
\be
\rho_I := e^{i H_0 t_i} \rho e^{- i H_0 t_i}\;.
\ee
However, due to the condition \eq{cond_rho} when applying \eq{chi_def} for $\chi(v)$, this implies that
\be\label{rho_I_rho}
\rho_I = \rho. 
\ee
In this work, we will restrict to the class of $\rho$ satisfying \eq{cond_rho}, thus \eq{rho_I_rho}. Examples of such $\rho$ can be energy eigenstates, thermal states, or sums of their direct products.

A reminder to readers is that we have suppressed, and will continue to suppress, the dependence on the spatial coordinates in the evolution operators. More formally, we should write the exponential factor as follows:
\be\label{global_x_0}
\int d^4x \lambda^{\pm}(x^{\mu};v) V_I(x^{\mu})
\ee
and accordingly, for the later expressions of $\chi(v)$ in terms of Green functions in real time. The dependence on spatial coordinates is important when considering the moving probes coupled to the environmental bath. Then, the probes' motions will also affect the work statistics. However, in this work, we will consider only static probes to simplify the calculations. Moreover, we will also assume the tuning of the coupling is ultra-local, i.e.,
\be\label{global_x}
\lambda^{\pm}(x^{\mu};v) = \lambda^{\pm}(t;v)  \delta(\vec{x}-\vec{x}_0)
\ee
with $\vec{x}$ the spatial coordinate vector. By this assumption, \eq{global_x_0} can be reduced to
\be
\int dt \; \lambda^{\pm}(t;v) V_I(t)
\ee
as adopted in this work with the implicit dependence on $\vec{x}_0$ omitted.

If we consider a cyclic process with zero coupling at the endpoints, i.e., $\lambda(t_i)=\lambda(t_f)=0$, then \eq{chiv_11} can be reduced to \cite{Ortega:2019etm}
\be\label{chiv_12}
\chi(v) = {\rm Tr} \Big[ {\cal T}_+\big[e^{-i \int_{t_i + v}^{t_f +v} \lambda(t-v) V_I(t) dt }  \big] \; \rho \; {\cal T}_-\big[e^{i \int^{t_f}_{t_i} \lambda(t) V_I(t) dt }  \big]   \Big] \;.
\ee
Otherwise, it seems that we cannot further reduce \eq{chiv_11} into a more compact form because the time-ordering operators do not form an abelian group, i.e., ${\cal T}_+ O_1 {\cal T}_+ O_2 \neq {\cal T}_+ O_1 O_2$ in general. However, in \eq{chiv_11}, the two time-ordering operators live on two non-overlapped real-time intervals so that we can formally merge them into a single time-ordering operator if we introduce the following time-piecewise couplings:
\begin{align}
    \label{lambda_p}
\lambda^+(t;v) :=&\, \lambda_i \Big[ \Theta(t-t_i) - \Theta(t-t_i-v) \Big] + \lambda(t-v) \Big[ \Theta(t-t_i-v) -\nonumber\\
& \Theta(t-t_f - v) \Big] \;, \\
\lambda^-(t;v) :=& \, \lambda(t) \Big[ \Theta(t-t_i) -\Theta(t-t_f) \Big] + \lambda_f \Big[\Theta(t-t_f) -\nonumber\\
&  \Theta(t-t_f-v) \Big] \;, \label{lambda_m}
\end{align}
where $\Theta(t)$ is the Heaviside step function. The superscript $\pm$ indicates the coupling constant on the forward and backward paths of Fig. \ref{contour}. Then, the WCF $\chi(v)$ can be put into the following compact form: 
\be
\chi(v) = {\rm Tr}\Big\{{\cal T}_+ \big[ e^{-i \int_{-\infty}^{\infty} dt \; \lambda^+(t;v) V_I(t)} \big] \; \rho \; {\cal T}_- \big[ e^{i \int_{-\infty}^{\infty} dt \; \lambda^-(t;v) V_I(t)} \big]
 \Big\} \;. \label{chi_realtime}
\ee
Note that the $v$-dependence of $\lambda^{\pm}(t;v)$'s is asymmetric because the insertions of $e^{-i H(0)v}$ and $e^{i H(T)} v$ in the real-time contour take the order of OTOC, which is asymmetric on the forward and backward paths. This asymmetry yields a fluctuation theorem, which implies a time arrow for the relaxation of the non-equilibrium process, manifesting the second law microscopically. The above  result was also proposed in \cite{fei2020nonequilibrium} in the following form:
\be\label{chi_C}
\chi(v)={\langle {\cal T}_{C} \big[ e^{-i \int_{C} dt\; \lambda_{C}(t) V_I(t)} \big] \rangle \over  \langle {\cal T}_{C} \big[ e^{-i \int_{C} dt\; \lambda(0) V_I(t)} \big] \rangle}
\ee
where $C$ is the extended real-time contour, as shown in Fig. \ref{contour}, $\cal T_{C}$ the time ordering operator and $\lambda_{C}$ the piecewise coupling function along $C$.

\section{Generalized Wick theorem and its application to derive nonperturbative WCF of pure thermal bath}\label{app_wick}

We have considered the work characteristic function  $\chi(v)$ of pure thermal bath up to ${\cal O}(\lambda^2)$. In this section, we will generalize it to all orders of $\lambda$, at least formally.

\subsection{Review on general Wick's theorem}
To proceed with the $\chi(v)$'s derivations, we should invoke the generalized Wick's theorem \cite{Diosi:2017kzh, Ferialdi:2021ywo}. Consider two different orderings $\mathbb{O}$ and $\mathbb{O}'$ defined for the corresponding sets of bases, namely, $\{\phi_a \}$ and $\{\psi_b \}$, as follows:
\bea
&& \mathbb{O}[\phi_1\cdots \phi_n] = (\pm 1)^P \phi_{p_1} \cdots \phi_{p_n}\;, 
\\
&& \mathbb{O}'[\psi_1\cdots \psi_n] = (\pm 1)^P \psi_{p_1} \cdots \psi_{p_n}
\eea
with $P$ the number of permutations bringing the order list $(1,\cdots,n)$ to the one $(p_1,\cdots,p_n)$. 
Then, the general Wick's theorem relating the orderings $\mathbb{O}$ and $\mathbb{O}'$ states that 
\be\label{Wick_thm}
\mathbb{O}\Big[ e^{\int  \lambda(x) \phi(x) d^4x } \Big] =e^{{1\over 2}\int\int  \lambda(x) \lambda(y) C(x,y) d^4x d^4y} \; \mathbb{O}'\Big[ e^{\int  \lambda(x) \phi(x) d^4x} \Big]
\ee
where
\be
C(x,y)=(\mathbb{O} - \mathbb{O}') \phi(x) \phi(y)
\ee
is a pure commutator and thus a c-number, as long as the two sets of operators $\phi_a$ and $\psi_b$ are linearly related,
\be\label{linear_phi_psi}
\phi_a = \mathcal{L}_{ab} \psi_b\;.
\ee
Since $C(x,y)$ is a c-number, so that for the state $\rho$ annihilated by the ordering $\mathbb{O}'$, the general Wick's theorem \eq{Wick_thm} implies
\begin{align}
    {\rm Tr}\Bigg[\mathbb{O}\Big[ e^{\int  \lambda(x) \phi(x) d^4x } \Big] \Bigg] 
    =&\, e^{{1\over 2}\int\int  \lambda(x) \lambda(y) C(x,y) d^4x d^4y} \nonumber\\
    =&\, e^{{1\over 2}\int\int  \lambda(x) \lambda(y) {\rm Tr}\big[\rho \mathbb{O}\phi(x)\phi(y)  \big] d^4x d^4y} \;.
\end{align}

 To demonstrate the general Wick's theorem by examples relating to the normal ordering $\mathcal{N}$, we mode-decompose the hermitian field operators $\phi_i:=\phi(x_i)$ as follows,
\be
\phi_i = \sum_k \sqrt{\mathcal{S}_k} \; \Big[ a_k e^{-i k.x_i} + a_k^{\dagger}e^{i k.x_i}\Big] \qquad {\rm with} \quad k.x_i= \omega_k t_i - \vec{k}\cdot \vec{x}_i\;,
\ee
where $\mathcal{S}_k$ is the spectral density and 
\be
[a_k, a^{\dagger}_p]_{\epsilon} =\delta_{k,p}\;, \qquad [a_k, a_p]_{\epsilon} = [a^{\dagger}_k, a^{\dagger}_p]_{\epsilon}=0
\ee
with $\epsilon=+1$ for bosonic operators and $\epsilon=-1$ for the fermionic ones, with the commutator defined by
\be
\big[A, B \big]_{\epsilon} := A B - \epsilon B A\;.
\ee
Note that the ground state $|0\rangle$ is defined by $a_k|0\rangle =0$, $\forall k$. Generalizing the above discussions to the non-hermitian field is straightforward and can be found in \cite{Evans:1996bha, Evans:1998yi}.

We would like to consider the normal ordering $\mathcal{N}_{\rho}$ for general state $\rho$, which is characterized by the number density $n_k$ with $k$ the momentum mode label, i.e.,
\be\label{rho_number}
{\rm Tr}[\rho a_k^{\dagger} a_p] = n_k \delta_{k,p}\;, \qquad  {\rm Tr}[\rho a_k a_p^{\dagger}] = (1+ \epsilon n_k) \delta_{k,p}\;.
\ee
For the ground state, $n_k=0$. On the other hand, for the thermal $\rho_{\beta}$, $n_k$ is the thermal weight of mode $k$, i.e., 
\be
n_k={1\over e^{\beta \omega_k} - \epsilon}
\ee
with $\omega_k$ the energy of the mode $k$. 

Postulating the normal ordering $\mathcal{N}_{\rho}$ is defined by
\be\label{Nrho_def}
\mathcal{N}_{\rho}[\phi_1 \phi_2]:= \phi_1^+ \phi_2^+ + \phi_1^- \phi_2^+ + \epsilon \phi_2^- \phi_1^+ + \phi_1^- \phi_2^-
\ee
with
\bea\label{phi_p_def}
\phi^+_i &=& \sum_k \sqrt{\mathcal{S}_k} \; \Big[ (1- f_k) a_k e^{-i k.x_i} + g_k a_k^{\dagger}e^{i k.x_i}\Big]\;,
\\ \label{phi_m_def}
\phi^-_i &=& \sum_k \sqrt{\mathcal{S}_k} \; \Big[ f_k a_k e^{-i k.x_i} + (1- g_k) a_k^{\dagger}e^{i k.x_i}\Big]\;.
\eea
Note that
\be
\phi_i = \phi^+_i + \phi^-_i\;.
\ee
The coefficients $f_k$ and $g_k$ shall be determined by the following constraints
\be\label{Nrho_c}
0={\rm Tr}\big[\rho \mathcal{N}_{\rho}[\phi_1 \phi_2] \big] = {\rm Tr}\big[\rho \mathcal{N}_{\rho}[\phi_2 \phi_1] \big]
\ee
which respectively yield
\bea\label{fg_0}
 f_k + g_k &=& f_k g_k -\epsilon n_k\;,
\\ \label{fg_1}
f_k g_k &=& -\epsilon n_k\;.
\eea
They further yield
\be \label{fg_2}
(1-f_k)(1-g_k) = 1 + \epsilon n_k\;. 
\ee

Let's apply the general Wick's theorem relating the normal ordering $\mathcal{N}_{\rho}$ to calculate the real-time Green functions of state $\rho$. First, consider the non-ordered Green functions $iG^{-+}$ or $iG^{+-}$, which can be expressed as 
\be
{\rm Tr}\Big[\rho \phi_1 \phi_2  \Big] =\sum_k \mathcal{S}_k \Big[ n_k e^{i k.(x_1-x_2)} + (1+\epsilon n_k) e^{-i k.(x_1-x_2)} \Big] 
\ee
with $\rho$ obeying the condition \eq{rho_number} and \eq{Nrho_c}. We shall show
\be\label{Gmp_Wick}
{\rm Tr}\Big[\rho \phi_1 \phi_2  \Big] = \phi_1 \phi_2 - \mathcal{N}_{\rho}[\phi_1 \phi_2] = \big[\phi^+_1, \phi^-_2 \big]_{\epsilon}
\ee
The second equality of \eq{Gmp_Wick} is obtained by straightforwardly applying \eq{Nrho_def}, \eq{phi_p_def} and \eq{phi_m_def}. After a straightforward calculation, one can show that \eq{Gmp_Wick} holds by using \eq{fg_1} and \eq{fg_2}. 

Next, we consider the time-ordered Green function, e.g.,
\be\label{GR_Wick_def}
{\rm Tr}\Big[\rho  \mathcal{T}_+ \phi_1 \phi_2  \Big] =\Theta(t_1-t_2) {\rm Tr}\Big[\rho \phi_1 \phi_2  \Big] + \epsilon \Theta(t_2-t_1) {\rm Tr}\Big[\rho \phi_2 \phi_1  \Big]\;.
\ee
We shall show that
\begin{align}
    \label{GR_Wick_qed}
&{\rm Tr}\Big[\rho  \mathcal{T}_+ \phi_1 \phi_2  \Big]\nonumber\\
=&\,  \mathcal{T}_+ \Big[ \phi_1 \phi_2  \Big] - \mathcal{N}_{\rho}[\phi_1 \phi_2] 
\\ \label{GR_Wick}
=&\, \Theta(t_1-t_2) \big[\phi_1^+, \phi_2^- \big]_{\epsilon} -\Theta(t_2-t_1) \Big( \big[\phi_1^+, \phi_2^+ \big]_{\epsilon} + \big[\phi_1^-, \phi_2^+ \big]_{\epsilon} \nonumber\\
&\, + \big[\phi_1^-, \phi_2^- \big]_{\epsilon} \Big)\;.
\end{align}

Again, we have used \eq{Nrho_def}, \eq{phi_p_def}, and \eq{phi_m_def} to arrive at the second equality. Note that the first term of \eq{GR_Wick} agrees with the first term of \eq{GR_Wick_def} using \eq{Gmp_Wick}. The three other commutators of \eq{GR_Wick} can be calculated straightforwardly, and the results are 
\begin{align}
    &\big[\phi_1^-, \phi_2^+ \big]_{\epsilon}\nonumber\\
    =&\, \sum_k \mathcal{S}_k \Big[ f_k g_k e^{-i k.(x_1-x_2) } -\sigma (1-f_k)(1-g_k) e^{i k.(x_1-x_2) } \Big]\;, 
\\
&\big[\phi_1^+, \phi_2^+ \big]_{\epsilon} + \big[\phi_1^-, \phi_2^- \big]_{\epsilon} \nonumber\\
=&\, \sum_k \mathcal{S}_k \Big(e^{-i k.(x_1-x_2) } -e^{i k.(x_1-x_2) } \Big) \Big[ (1-f_k) g_k+ f_k (1-g_k)  \Big]\;.
\end{align}

Applying \eq{fg_0}, \eq{fg_1} and \eq{fg_2}, we can show that
\bea
\big[\phi_1^-, \phi_2^+ \big]_{\epsilon} &=& -\epsilon {\rm Tr}\Big[\rho \phi_2 \phi_1  \Big]\;, 
\\
\big[\phi_1^+, \phi_2^+ \big]_{\epsilon} + \big[\phi_1^-, \phi_2^- \big]_{\epsilon} &=& 0\;.
\eea
Plugging the results into \eq{GR_Wick}, we verify \eq{GR_Wick_qed} holds. Similar consideration should work for reverse time order $\mathcal{T}_-$.

In summary, the relations \eq{Gmp_Wick} and \eq{GR_Wick_qed} imply that the linear transformation \eq{linear_phi_psi} is the Bogoliubov transformation that relates the ground state to the thermal state in the real-time ordered path approach. However, in the imaginary-time approach, no sign of a thermal Bogoliubov transformation was shown \cite{Blasone:1997nt}. Moreover, for state $\rho$ and the associated normal ordering $\mathcal{N}_{\rho}$ satisfying \eq{Nrho_c}, the general Wick theorem implies that
\be\label{merge_1}
\mathbb{O}\Big[\mathcal{F}(\phi)\Big] = e^{{1\over 2}\int d^4x \int d^4y \; {\rm Tr}\big[\rho \mathbb{O}\phi(x)\phi(y)  \big] {\delta^2 \over \delta \phi(x) \delta \phi(y)} }  \mathcal{N}_{\rho}\Big[\mathcal{F}(\phi)\Big]
\ee
and a corollary \cite{Polchinski:1998rq}
\begin{align}
    \label{merge_2}
&\mathcal{N}_{\rho}\Big[\mathcal{F}(\phi)\Big] \mathcal{N}_{\rho}\Big[\mathcal{G}(\psi)\Big]\nonumber\\
=&\, e^{{1\over 2}\int d^4x \int d^4y \; {\rm Tr}\big[\rho \phi(x)\psi(y)  \big] {\delta^2 \over \delta \phi(x) \delta \psi(y)} }\mathcal{N}_{\rho}\Big[\mathcal{F}(\phi) \mathcal{G}(\psi)\Big]
\end{align}
\
In the next subsection, we will apply \eq{merge_1} and \eq{merge_2} to derive $\chi(v)$ to all orders.

\subsection{Characteristic function from general Wick's theorem}

We now apply \eq{merge_1} and \eq{merge_2} to calculate  \eq{chi_realtime}, i.e., 
\be\label{chi_f_Opm}
\chi(v) = {\rm Tr}\Big\{{\cal T}_+ e^{-i O^+_v }  \; \rho \; {\cal T}_- e^{i O^-_v } \Big\} 
\ee
with 
\be
O^{\pm}_v := \int_{\infty}^{-\infty} dt \; \lambda^{\pm}(t; v) V_I(t)\;.
\ee
We start the example for $V$ as the bosonic operator $O$ in the initial thermal state $\rho_{\beta}$. We denote the normal ordering by $\mathcal{N}_{\beta}$ for simplicity. Then, applying \eq{merge_1} and \eq{merge_2},
we can reduce \eq{chi_f_Opm} into
\begin{align}
    \label{chi_all_thermal}
\chi(v)=&\,e^{-{1\over 2}  \big( \langle O_v^+ O_v^+ \rangle_{\beta} + \langle O_v^- O_v^- \rangle_{\beta} \big) } \langle \mathcal{N}_{\beta}\big[e^{i O_v^-}\big] \mathcal{N}_{\beta}\big[e^{-i O_v^+}\big] \rangle_{\beta} \nonumber\\
=&\, e^{-{1\over 2}\big( \langle O_v^+ O_v^+ \rangle_{\beta} + \langle O_v^- O_v^- \rangle_{\beta} \big) + \langle O_v^- O_v^+ \rangle_{\beta} } \langle \mathcal{N}_{\beta}\big[ e^{i (O_v^- - O_v^+)}\big] \rangle_{\beta} \nonumber\\
=&\, e^{\chi^{(2)}(v) -1}
\end{align}

where
\begin{align}
    \langle O_v^+ O_v^+ \rangle_{\beta} =&\, \int_{-\infty}^{\infty} dt \int_{-\infty}^{\infty} dt' \lambda^+(t;v) \lambda^+(t';v) \; iG^{++}_{\beta}(t,t')\;,
\\
\langle O_v^- O_v^- \rangle_{\beta} =&\, \int_{-\infty}^{\infty} dt \int_{-\infty}^{\infty} dt' \lambda^-(t;v) \lambda^-(t';v) \; iG^{--}_{\beta}(t,t')\;,
\\
\langle O_v^- O_v^+ \rangle_{\beta} =&\, \int_{-\infty}^{\infty} dt \int_{-\infty}^{\infty} dt' \lambda^-(t;v) \lambda^+(t';v) \; iG^{-+}_{\beta}(t,t') \nonumber\\
=&\, \int_{-\infty}^{\infty} dt \int_{-\infty}^{\infty} dt' \lambda^+(t;v) \lambda^-(t';v) \; iG^{+-}_{\beta}(t',t) \;,
\end{align}

$\chi^{(2)}(v)$ is the ${\cal O}(\lambda^2)$ result given in \eq{chiv_omega_2}, and $iG^{\pm\pm}_{\beta}(t)$ are the real-time thermal Green functions of $O(t)$. We have used the constraint \eq{Nrho_c} on $\mathcal{N}_{\beta}$ and $\rho_{\beta}$ and the same procedure of reducing \eq{chiv_omega_1_m} to \eq{chiv_omega_2} to arrive at the last equality of \eq{chi_all_thermal}. Since $\chi^{(2)}(v)=1$ for $\rho=\rho_{\beta}$, this implied that $\chi(v)=1$ for the thermal states. Therefore, we have shown that the thermal states are non-perturbatively passive.

Now, we may apply the general Wick's theorem to calculate $\chi(v)$ to the cases with qubit coupling, in which the interactions are a composite operator of \eq{qubit_V}. The quadratic nature of such interactions breaks the linear relation \eq{linear_phi_psi} among the bases of different orderings. Thus, we cannot apply general Wick's theorem or the merging formulas \eq{merge_1} and \eq{merge_2} directly. On the other hand, we can keep the qubit operators as the coefficients of the q number and introduce only the normal ordering $\mathcal{N}_{\beta}$ for the thermal environment operators. Applying the merging formulas \eq{merge_1} and \eq{merge_2} to $\mathcal{N}_{\beta}$, we can formally arrive at a succinct formula for $\chi(v)$ with the initial state $\rho=\rho_{\rm SQ}\otimes \rho_{\beta}$. Take the case of a spin-qubit as an example. Adopting the notations of spin operator $\sigma_x(t)$ defined in \eq{sigma_x_t} and keeping the positions of the q-number operator $\sigma_x$ intact, the result can be written as   
\begin{align}
    \label{chi_v_elegant}
\chi(v)=&\,{\rm Tr} \Big[ e^{-{1\over 2}\sum_{a,b=\pm 1} \epsilon^{ab} \langle O_v^a O_v^b \rangle_{\beta}}  \mathcal{N}_{\beta}\big[ e^{i (O_v^- - O_v^+)}\big] \; \rho_{\rm SQ}\otimes \rho_{\beta} \Big] \nonumber\\
=&\, {\rm Tr} \Big[  e^{-{1\over 2}\sum_{a,b=\pm 1} \epsilon^{ab} \langle O_v^a O_v^b \rangle_{\beta}}   \rho_{\rm SQ} \Big]
\end{align}
with
\begin{align}
     \langle O_v^+ O_v^+ \rangle_{\beta} =&\,\int_{-\infty}^{\infty}dt \int_{-\infty}^{\infty}dt' \lambda^+(t;v) \lambda^+(t';v)\; iG^{++}_{\beta}(t,t') \nonumber\\
     &\,\times\big[\sigma_x(t) \sigma_x(t') \; \boldsymbol{\cdot}\big]\;,
\\
 \langle O_v^- O_v^- \rangle_{\beta} =&\,\int_{-\infty}^{\infty}dt \int_{-\infty}^{\infty}dt' \lambda^-(t;v) \lambda^-(t';v)\; iG^{--}_{\beta}(t,t') \nonumber\\
 &\,\times\big[ \boldsymbol{\cdot} \sigma_x(t) \sigma_x(t') \big] \;, 
\\
 \langle O_v^- O_v^+ \rangle_{\beta} =&\,\int_{-\infty}^{\infty}dt \int_{-\infty}^{\infty}dt' \lambda^-(t;v) \lambda^+(t';v)\; iG^{-+}_{\beta}(t,t') \nonumber\\
 &\,\times\big[ \sigma_x(t)\boldsymbol{\cdot} \sigma_x(t') \big] \;,
\\
\langle O_v^+ O_v^- \rangle_{\beta} =&\,\int_{-\infty}^{\infty}dt \int_{-\infty}^{\infty}dt' \lambda^+(t;v) \lambda^-(t';v)\; iG^{+-}_{\beta}(t,t') \nonumber\\
&\,\times\big[ \sigma_x(t') \boldsymbol{\cdot} \sigma_x(t) \big] \;,
\end{align}

where $\boldsymbol{\cdot}$ denotes an operator such as $\rho_{\rm SQ}$ on which $\sigma_x(t)$ or $\sigma_x(t')$ will act, and their relative ordering is required by the corresponding time ordering. We have used the constraint \eq{Nrho_c} in $\mathcal{N}_{\beta}$ and $\rho_{\beta}$ to arrive at the last equality of \eq{chi_v_elegant}. 
A similar expression for $\chi(v)$ can be obtained for the case of the fermion qubit. Although the expression of \eq{chi_v_elegant} is elegant, it is still difficult to further simplify it to a closed form for all orders. Thus, in practice, we will still expand \eq{chi_v_elegant} to ${\cal O}(\lambda^2)$, which is nothing but the $\chi^{(2)}(v)$ we have studied in the previous sections.

\section{Derivations of work statistics with spin qubit coupling to thermal states}\label{app_C}

This section considers the work statistics and extraction from thermal states by a spin qubit. The spin qubit we consider is a two-level system with an energy gap $\Omega$ between ground state $|0\rangle$ and excited state $|1\rangle$, i.e., $H_0^{\rm SQ} = \Omega |1\rangle \langle 1|$, and consider the work statistics for an impulse process with the Hamiltonian and the initial state chosen to be 
\be\label{H_spin_qubit}
H(t)=H_0^{\rm TB} + H_0^{\rm SQ} + \lambda(t) V \quad {\rm with} \quad V= \sigma_x \otimes O\;, \quad 
\rho = \rho_{\rm SQ} \otimes \rho_{\beta}
\ee
where the spin operator $\sigma_x :=|0\rangle \langle 1| + |1\rangle \langle 0|$ and $\rho_{\rm SQ}$ should be chosen to be commuting with $H_0^{\rm SQ}$ so that 
\be
\rho_{\rm SQ} = p |0\rangle \langle 0| + (1-p) |1\rangle \langle 1| 
\ee
with $0\le p \le 1$. Note that $\rho$ chosen here satisfies the condition \eq{cond_rho}, or equivalently \eq{rho_I_rho}.

The counterpart of $V$ in the interaction picture is 
\be
V_I(t) = \sigma_x(t) \otimes O(t) 
\ee
with
\be\label{sigma_x_t}
\quad \sigma_x(t):=e^{- i \Omega t} |0\rangle \langle 1| + e^{ i \Omega t} |1\rangle \langle 0| \quad  {\rm and} \quad O(t):=e^{i H_0^{\rm TB} t} O e^{-i H_0^{\rm TB} t} \;.
\ee

To evaluate $\chi^{(2)}(v)$ of \eq{chiv_omega_1_m}, we first evaluate the corresponding (Fourier transform of) real-time Green functions for the total system, which can be expressed in terms of the product of the component ones as follows:
\begin{align}
    \label{G_spin_tot_1} 
i \tilde{G}^{-+}(\omega) =&\, \int_{-\infty}^{\infty} dt \; e^{i\omega t} \; {\rm Tr}\Big(\rho V_I(t) V_I(0) \Big)\nonumber\\
=&\, \int_{-\infty}^{\infty} dt \; e^{i\omega t} iG^{-+}_{\sigma}(t) \; iG^{-+}_{\beta}(t) \; \;,  \\
i \tilde{G}^{+-}(\omega) =&\, \int_{-\infty}^{\infty} dt \; e^{i\omega t} \; {\rm Tr}\Big(\rho V_I(0) V_I(t) \Big)\nonumber\\
=&\, \int_{-\infty}^{\infty} dt \; e^{i\omega t} iG^{+-}_{\sigma}(t) \; iG^{+-}_{\beta}(t) \; \;. \label{G_spin_tot_2}
\end{align}
with 
\be
iG^{-+}_{\sigma}(t) :={\rm Tr}\Big( \rho_{\rm SQ} \sigma_x(t) \sigma_x(0) \Big) =  p e^{- i \Omega t}  + (1-p)  e^{ i \Omega t} \label{G_sigma1}  
\ee
and
\be
iG^{+-}_{\sigma}(t) :={\rm Tr}\Big( \rho_{\rm SQ} \sigma_x(0) \sigma_x(t) \Big) = iG^{-+}_{\sigma}(-t)  \label{G_sigma2}
\ee
and the thermal Green functions $iG^{-+}_{\beta}(t)$ and  $iG^{+-}_{\beta}(t)$ satisfying the KMS condition \eq{KMS_m}.

Applying to \eq{G_spin_tot_1} and \eq{G_spin_tot_2} the Fourier transform of \eq{G_sigma1} and \eq{G_sigma2}
\be
i\tilde{G}^{-+}_{\sigma}(\omega) = p 2 \pi \delta(\omega - \Omega) +  (1-p) 2\pi \delta(\omega + \Omega)
\ee
and 
\be
i\tilde{G}^{+-}_{\sigma}(\omega) = i\tilde{G}^{-+}_{\sigma}(-\omega)\;,
\ee
we obtain
\begin{align}
    \label{UDW_Gmp}
 i \tilde{G}^{-+}(\omega) =&\, \int_{-\infty}^{\infty} {d\omega' \over 2\pi}  i\tilde{G}^{-+}_{\sigma}(\omega -\omega') S^O_{\beta}(\omega') \nonumber\\
 =&\, p S^O_{\beta}(\omega - \Omega) + (1-p) S^O_{\beta}(\omega + \Omega) \;, \\
 i \tilde{G}^{+-}(\omega) =&\, \int_{-\infty}^{\infty} {d\omega' \over 2\pi}  i\tilde{G}^{-+}_{\sigma}(\omega - \omega') e^{-\beta \omega'} S^O_{\beta}(\omega') \nonumber\\
 =&\, e^{-\beta \omega} \Big[ p e^{-\beta \Omega} S^O_{\beta}(\omega + \Omega) + (1-p)  e^{\beta \Omega}  S^O_{\beta}(\omega - \Omega) \Big]\;, \label{UDW_Gpm}
\end{align} 
with $S^O_{\beta}(\omega)$ defined in \eq{KMS_m}.
Note that by setting $\Omega=0$, $i\tilde{G}^{-+}(\omega)$ reduces to $S^O_{\beta}(\omega)$ and $i\tilde{G}^{+-}(\omega)$ to $e^{-\beta \omega} S^O_{\beta}(\omega)$. The corresponding $P^{(2)}(W)$'s are the same as those without adding the qubit.  In terms of Pauli-Jordan spectral function, \eq{UDW_Gmp} and \eq{UDW_Gpm} can be turned into
\begin{align}
    \label{UDW_Gmp_1}
i \tilde{G}^{-+}(\omega) =&\, p e^{\beta(\omega-\Omega)} \; n_{\rm BE}(\omega - \Omega) S^O(\omega - \Omega) + \nonumber\\
&(1-p) e^{\beta(\omega+\Omega)} \;  n_{
\rm BE}(\omega + \Omega) S^O(\omega + \Omega)  \\ \label{UDW_Gpm_1}
i \tilde{G}^{+-}(\omega) =&   (1-p)  \; n_{\rm BE}(\omega - \Omega) S^O(\omega - \Omega) +\nonumber\\
&p \;  n_{
\rm BE}(\omega + \Omega) S^O(\omega + \Omega)
\end{align} 
and thus
\begin{widetext}
    \begin{align}
    \label{main_1}
i \tilde{G}^{-+}(\omega) \pm i \tilde{G}^{+-}(\omega)  =&\, \big[p \pm (1-p)e^{-\beta(\omega-\Omega)} \big] \; S^O_{\beta}(\omega - \Omega) + \big[1-p \pm p e^{-\beta(\omega+\Omega)} \big]\; S^O_{\beta}(\omega + \Omega)   \;, \\
=&\, \big[p e^{\beta(\omega-\Omega)} \pm (1-p) \big] \; n_{\rm BE}(\omega - \Omega) S^O(\omega - \Omega) + \big[(1-p) e^{\beta(\omega+\Omega)} \pm p  \big]\;  n_{
\rm BE}(\omega + \Omega) S^O(\omega + \Omega) \;. \qquad 
\end{align}
\end{widetext}

We can then plug \eq{UDW_Gmp} and \eq{UDW_Gpm} into \eq{chiv_omega_1_m} to obtain the explicit expression of $\chi^{(2)}(v)$. We now study the implications of the result for the work statistics and the work extraction by the spin qubit. First, consider the violation of Jarzynski's equality by \eq{chi_beta2_G} to obtain
\begin{widetext}
    \begin{equation}
    \overline{e^{-\beta W}}\Big\vert_{{\cal O}(\lambda^2)}=\chi^{(2)}(i\beta)=1-{1\over 2} \int_{\lambda} \big(1-e^{-\beta \omega} \big) \Big\{ \big[p - (1-p) e^{\beta \Omega} \big] S_{\beta}^O(\omega - \Omega) + \big[ 1- p - p e^{-\beta \Omega} \big] S_{\beta}^O(\omega+ \Omega) \Big\}\;.
\label{chi_2_sq}
\end{equation}
\end{widetext}
In the above, we have used the short-hand notation $\int_{\lambda}$ defined in \eq{short-h}, and will do the same below.

For $\beta \Omega \ll 1$, we can expand the above to ${\cal O}\big((\beta \Omega)^2\big)$, and the result is 
\begin{align}
    \chi^{(2)}(i\beta)=&\,1 + \big(2p-1 - {1\over 2} \beta \Omega \big)\; \beta \Omega \; \int_{\lambda} \nonumber\\
    &\big(1-e^{-\beta \omega} \big) \Big[{1\over \beta}{d S_{\beta}^O \over d\omega}(\omega) -{1\over 2} S^O_{\beta}(\omega) \Big]\;.
\end{align}

Next, we can evaluate the work statistics $P^{(2)}(W)$ and the average work done $\overline{W}^{(2)}$ by using \eq{p_W_1_m}, \eq{m_p0} and \eq{W_ext_f}, respectively, with the help of \eq{UDW_Gmp}, \eq{UDW_Gpm}.
For $p= 0, 1$,
\begin{align}
    &P^{(2)}(W)|_{W\ne 0}\begin{cases}
        p=1\\
        p=0
    \end{cases}\nonumber\\
    =&\,{1\over 4\pi} |\tilde{\lambda}(W)|^2  \begin{cases}
    \big[ S^O_{\beta}(W - \Omega) +  e^{\beta (W - \Omega)} S^O_{\beta}(-W + \Omega)  \big]\;, \\
    \big[ S^O_{\beta}(W + \Omega)  + e^{\beta (W + \Omega)} S^O_{\beta}(-W - \Omega)  \big]\;, 
  \end{cases}   \qquad  \label{P_2W_10}
\end{align}
and
\begin{align}
    &\overline{W}^{(2)}\begin{cases}
        p=1\\
        p=0
    \end{cases} \nonumber\\
    =&\, {1\over 2} \int_{\lambda} \omega\;   \begin{cases}
    \big[ S^O_{\beta}(\omega - \Omega)  - e^{-\beta (\omega + \Omega)} S^O_{\beta}(\omega + \Omega)  \big]\;, \\
    \big[ S^O_{\beta}(\omega + \Omega)  - e^{-\beta (\omega - \Omega)} S^O_{\beta}(\omega - \Omega)  \big]\;.
  \end{cases}  \qquad \label{Wbar_1}
\end{align}

For $p=1/2$, 
\bea
 P^{(2)}(W)|_{p={1\over 2}}&=& {1\over 2}\Big[P^{(2)}(W)|_{p=0} + P^{(2)}(W)|_{p=1} \Big]\;, \\
 \overline{W}^{(2)}\big\vert_{p={1\over 2}}&=&{1\over 2}\Big[\overline{W}^{(2)}\big\vert_{p=0} + \overline{W}^{(2)}\big\vert_{p=1} \Big]\;.
\eea
From \eq{Wbar_1}, we find that $\overline{W}^{(2)}\big\vert_{p=1, \beta=0} = -\overline{W}^{(2)}\big\vert_{p=0, \beta=0}<0$, so $\overline{W}^{(2)}\big\vert_{p={1\over 2}, \beta=0}=0$. Thus, it is possible to extract work using the above setup, at least in a high-temperature environment with no population inversion for the qubit system. On the other hand, $\overline{W}^{(2)}|_{\beta=\infty}>0$ if $S^O_{\beta=\infty}(\omega)$ is a monotone of $\omega$. Otherwise, it could be negative unless $p=0$. All of the above are consistent with expectations.

\section{Derivations of work statistics with fermion or topological qubit coupling to thermal states}\label{app_D}

In this section, we will consider the work statistics for the system with a fermion qubit coupled to the thermal environment. This fermion qubit can be a Dirac or a topological one made of two Majorana zero modes (MZMs).  The following Hamiltonian dictates the fermion qubit we consider, 
\be
H_0^{\rm FQ}=\Omega d^{\dagger} d 
\ee
where $d$
and $d^{\dagger}$ are fermionic annihilation and creation operators obeying
$\{d,d^{\dagger} \}=1$ and $d^2=(d^{\dagger})^2=0$. Then, the Hamiltonian for the total system with the qubit coupled to the thermal environment is  
\be\label{H_fermion_qubit}
H=H_0^{\rm TB} + H_0^{\rm FQ} + \lambda(t) V \qquad {\rm with} \qquad  V= d^{\dagger} \Psi + \Psi^{\dagger} d 
\ee
where $\Psi$ is a fermion quasiparticle operator characterized by its Pauli-Jordan spectral function
\be\label{f_Hadamard}
S^{\Psi}(\omega):=\int_{-\infty}^{\infty} dt \; e^{i\omega t} \big\{\Psi^{\dagger}(t), \Psi(0) \big\}\;.
\ee
The stability of the thermal states is ensured if $S^{\Psi}(\omega)$ is bounded below. 

We want to consider both the Dirac fermion qubit and the topological qubit of MZMs at the same time. Hence, we will rewrite the Dirac fermion into two MZMs to consider the work statistics.  Formally, we introduce two MZMs denoted by $\gamma_m$ with $m=1,2$, which satisfy the Clifford algebra $\{ \gamma_m, \gamma_n \}:=\gamma_m\gamma_n+\gamma_n\gamma_m=2\delta_{m,n}$, and $\gamma_m^{\dagger}=\gamma_m$. Then, the Dirac fermion can be decomposed into MZMs as follows \cite{kitaev2001unpaired, kitaev2010topological}:
\be\label{Kitaev}
d:={\gamma_1 + i \gamma_2 \over 2}, \quad d^{\dagger}={\gamma_1-i \gamma_2 \over 2}\;,\;
\ee
or
\be
\gamma_1= d^{\dagger} + d \;, \quad \gamma_2= i (d^{\dagger}-d).
\ee
Then, we can rewrite the Hamiltonians $H_0^{\rm FO}$ and $V$ into the following:
\begin{align}
    & H_0^{\rm FQ} =\, {\Omega \over 2} \Big(1+ i \gamma_1 \gamma_2 \Big)\;, \quad V = -i \sum_{m=1,2} \gamma_m O_m 
\end{align}
with
\begin{align}
    \quad  O_1:={i \over 2}\Big(\Psi - \Psi^{\dagger} \Big)\;, \quad O_2:= {1\over 2}\Big(\Psi + \Psi^{\dagger} \Big). \label{def_O_psi}
\end{align}

Note that $O_{1,2}$ are hermitian fermionic operators.

Then, we will consider the initial state given by
\be
\rho = \rho_{\rm FQ} \otimes \rho_{\beta}\quad {\rm with} \quad \rho_{\rm FQ}= p |0\rangle \langle 0| + (1-p) |1\rangle \langle 1| \;.
\ee
Here, we have chosen the representation for $\rho_{\rm FQ}$ by which $d^{
\dagger}=|1\rangle\langle 0|$ and $d=|0\rangle\langle 1|$.

When considering the topological qubit of two separated MZMs, the operators $O_1$ and $O_2$ are far separated. The causality then implies the cluster decomposition
\be\label{cluster_p}
\langle (O_1)^k  (O_2)^l \rangle_{\beta} \simeq \langle (O_1)^k \rangle_{\beta} \langle (O_2)^l \rangle_{\beta} \delta_{k, 2 {\bf N}} \delta_{l, 2 {\bf N}}
\ee
where the Kornecker deltas $\delta_{k, 2 {\bf N}}$ and $\delta_{l, 2 {\bf N}}$ ensure that only correlators with even fermion parity are non-zero.

Using the fact $\Big({1+ i \gamma_1 \gamma_2 \over 2} \Big)^2 = {1+ i \gamma_1 \gamma_2 \over 2}$, we can derive
\be
\gamma_1(t):=e^{i H_0^{\rm FQ}t} \gamma_1 e^{-i H_0^{\rm FQ} t} = \gamma_1 \cos\Omega t + \gamma_2 \sin\Omega t\;
\ee
and
\be
\gamma_2(t):=e^{i H_0^{\rm FQ}t} \gamma_2 e^{-i H_0^{\rm FQ} t} = - \gamma_1 \sin\Omega t + \gamma_2 \cos\Omega t,
\ee
so that
\be\label{V_I_fermi}
V_I(t) := e^{i (H_0^{\rm TB} + H_0^{\rm FQ}) t} V e^{- i (H_0^{\rm TB} + H_0^{\rm FQ}) t} = -i \sum_{m=1,2} \gamma_m(t) O_m(t)\;.
\ee
Note that $\{\gamma_m(t), O_n(t) \}=0$.

\subsection{Work statistics with fermion qubit coupling to thermal bath}

To calculate $\chi^{(2)}(v)$ by \eq{chiv_omega_1_m} of the total system, including the qubit and the environment, we shall first evaluate the corresponding real-time Green functions for the composite operator $V_I(t)$ of \eq{V_I_fermi}, which is the sum of the direct products of two fermion operators. Let us start with $iG^{-+}(t)$ by expressing it in terms of component Green functions of the component fermionic operators, i.e.,
\be
iG^{-+}(t) = {\rm  Tr}\Big(\rho V_I(t) V_I(0)  \Big) = \sum_{m, n =1,2} iG^{-+}_{\gamma, mn}(t)\; iG^{-+}_{\beta,mn}(t) \label{iG_mp_tot}
\ee
with
\be
iG^{-+}_{\gamma, mn}(t) := {\rm Tr}\Big(\rho_{\rm FQ} \gamma_m(t) \gamma_n(0) \Big)\;, \quad iG^{-+}_{\beta,mn} := \langle  O_m(t) O_n(0)  \rangle_{\beta}
\label{G-+s}
\ee
where $iG^{-+}_{\gamma,mn}(t)$'s are the Green functions for $\gamma_m$ operators, and $iG^{-+}_{\beta,mn}$ for $O_m$ of the thermal environment.  Similarly, 
\be
iG^{+-}(t) = {\rm  Tr}\Big(\rho V_I(0) V_I(t)  \Big) = \sum_{m, n =1,2} iG^{+-}_{\gamma, mn}(t)\; iG^{+-}_{\beta,mn} \label{iG_pm_tot}(t)
\ee
with
\be
iG^{+-}_{\gamma, mn}(t) := - {\rm Tr}\Big(\rho_{\rm FQ} \gamma_n(0) \gamma_m(t) \Big)
\label{G+-s}
\ee
and 
\be
iG^{+-}_{\beta,mn}(t) := - \langle  O_n(0) O_m(t)  \rangle_{\beta}.
\ee
In the above, we have used the standard definitions of the real-time Green functions for either bosonic or fermionic operators as follows:
\bea
iG^{++}_{\rho,\epsilon}(t) &=& \Theta(t) \; iG^{-+}_{\rho,\epsilon}(t) +  \Theta(-t)\; iG^{+-}_{\rho,\epsilon}(t)\;, 
\label{G++g}
\\
iG^{--}_{\rho,\epsilon}(t) &=& \Theta(-t)\; iG^{-+}_{\rho,\epsilon}(t) +   \Theta(t)\; iG^{+-}_{\rho,\epsilon}(t)
\eea
where 
\be
iG^{-+}_{\rho,\epsilon}(t) = {\rm Tr}\big[ \rho A(t) B(0) \big]\;, \quad iG^{+-}_{\epsilon}(t)=\epsilon \; {\rm Tr}\big[ \rho B(0) A(t) \big]
\label{Gpmmpg}
\ee
with $\epsilon=+1$ for bosonic fields and $\epsilon=-1$ for fermionic fields. By construction,
\be
iG^{++}_{\rho,\epsilon}(t)+iG^{--}_{\rho,\epsilon}(t)= iG^{-+}_{\rho,\epsilon}(t) +  iG^{+-}_{\rho,\epsilon}(t)\;. 
\ee
Moreover, for the thermal vacuum state, i.e., $\rho=\rho_{\beta}$, one shall also impose the KMS condition
\be
iG^{+-}_{\beta,\epsilon}(t) = \epsilon\; iG^{-+}_{\beta, \epsilon}(t-i\beta)\;. \label{KMS_gen}
\ee

Thus, for the current case,
\be
iG^{+-}_{\gamma, mn}(t) = -iG^{-+}_{\gamma, mn}(-t) \;, \quad iG^{+-}_{\beta, mn}(t) = - iG^{-+}_{\beta, mn}(t-i\beta)\;. \label{KMS_FQ}
\ee
Besides, it is easy to see that
\begin{align}
    iG^{++}(t)=&\,\sum_{m,n=1,2} \Big[\Theta(t) iG^{-+}_{\gamma,mn}(t) iG^{-+}_{\beta,mn}(t) + \Theta(-t) iG^{+-}_{\gamma,mn}(t) iG^{+-}_{\beta,mn}(t)  \Big] \nonumber\\
    =&\, \sum_{m,n=1,2} iG^{++}_{\gamma,mn}(t) iG^{++}_{\beta,mn}(t) \;.
\end{align}

by noting that 
\begin{align}
    &iG^{++}_{\gamma,mn}(t) iG^{++}_{\beta,mn}(t) \nonumber\\
    = &\, \Big[\Theta(t) iG^{-+}_{\gamma,mn}(t) + \Theta(-t) iG^{+-}_{\gamma,mn}(t) \Big] \Big[\Theta(t) iG^{-+}_{\beta,mn}(t) +\nonumber\\
    &\,\Theta(-t) iG^{+-}_{\beta,mn}(t) \Big] \nonumber \\
    = &\,  \Theta(t) iG^{-+}_{\gamma,mn}(t) iG^{-+}_{\beta,mn}(t) + \Theta(-t) iG^{+-}_{\gamma,mn}(t) iG^{+-}_{\beta,mn}(t),
\end{align}
where $\Theta(t)\Theta(-t) = 0$ is used to arrive the second line.

To further simplify \eq{iG_mp_tot} and \eq{iG_pm_tot}, we note that
\begin{align}\label{G_gamma_t}
    \begin{cases}
iG^{-+}_{\gamma,11}(t) =\, iG^{-+}_{\gamma,22}(t) = p e^{-i \Omega t} + (1-p) e^{i \Omega t}, \\
iG^{-+}_{\gamma,12}(t) = - iG^{-+}_{\gamma,21}(t) =i \big[ p e^{-i \Omega t} - (1-p) e^{i \Omega t}   \big]\;.
\end{cases}
\end{align}

Applying \eq{G_gamma_t} to \eq{iG_mp_tot}, we obtain 
\begin{align}
    \label{Gmp_FQ_t}
&iG^{-+}(t)\nonumber\\
=&\, iG^{-+}_{\gamma,11}(t) \Big(iG^{-+}_{\beta,11}(t) + iG^{-+}_{\beta,22}(t)  \Big) + G^{-+}_{\gamma,12}(t) \Big( G^{-+}_{\beta,21}(t) - \nonumber\\
&\, G^{-+}_{\beta,12}(t)  \Big)
\\
=&\, p e^{- i\Omega t} \langle  \Psi^{\dagger}(t) \Psi(0) 
 \rangle_{\beta} + (1-p) e^{i \Omega t} \langle  \Psi(t) \Psi^{\dagger}(0) \rangle_{\beta} 
\end{align}
where we have used the relations
\begin{align}
    iG^{-+}_{\beta,11}(t) + iG^{-+}_{\beta,22}(t)\nonumber
    =&\, \sum_{m=1,2} \langle O_m(t) O_m(0) \rangle_{\beta} \nonumber\\
    =&\,  {1\over 2} \Big(  \langle  \Psi^{\dagger}(t) \Psi(0) 
 \rangle_{\beta} + \langle  \Psi(t) \Psi^{\dagger}(0) \rangle_{\beta} \Big)\;\\
G^{-+}_{\beta,21}(t) - G^{-+}_{\beta,12}(t)\nonumber
=&\, i\Big( \langle O_1(t) O_2(0) \rangle_{\beta} - \langle O_2(t) O_1(0) \rangle_{\beta} \Big)  \nonumber\\
=&\,  {1\over 2} \Big( \langle  \Psi^{\dagger}(t) \Psi(0) 
 \rangle_{\beta} -\langle  \Psi(t) \Psi^{\dagger}(0) \rangle_{\beta} \Big)\;. \label{iGmp_t_f}
\end{align}

Similarly, applying \eq{G_gamma_t}, \eq{KMS_FQ} to \eq{iG_pm_tot}, we obtain
\begin{align}
    \label{Gpm_FQ_t}
&iG^{+-}(t)\nonumber\\
=&\, iG^{-+}_{\gamma,11}(-t) \Big(iG^{-+}_{\beta,11}(t-i\beta) + iG^{-+}_{\beta,22}(t-i\beta)  \Big) \nonumber\\
&+ G^{-+}_{\gamma,12}(-t) \Big( G^{-+}_{\beta,21}(t-i\beta) - G^{-+}_{\beta,12}(t-i\beta)  \Big)
\\
=&\,  p e^{ i\Omega t} \langle  \Psi^{\dagger}(t-i\beta) \Psi(0) \rangle_{\beta} + (1-p) e^{- i \Omega t} \langle  \Psi(t-i\beta) \Psi^{\dagger}(0) \rangle_{\beta}\;. \label{iGpm_t_f}
\end{align}

Define the Wightman spectral functions
\be\label{FS_1}
S^{\Psi^{\dagger}\Psi}_{\beta}(\omega) = \int_{-\infty}^{\infty} dt \; e^{ i \omega t} \; \langle  \Psi^{\dagger}(t) \Psi(0) 
 \rangle_{\beta}  
\ee
and
\be\label{FS_2}
S^{\Psi \Psi^{\dagger}}_{\beta}(\omega) = \int_{-\infty}^{\infty} dt \; e^{ i \omega t} \; \langle \Psi(t) \Psi^{\dagger}(0)  
 \rangle_{\beta}
\ee
which are related to the Pauli-Jordan spectral function $S^{\Psi}(\omega)$ defined in \eq{f_Hadamard} by
\be
S^{\Psi}(\omega)= S^{\Psi^{\dagger}\Psi}_{\beta}(\omega) + S^{\Psi \Psi^{\dagger}}_{\beta}(\omega)  \;.
\ee
Using \eq{FS_1} and \eq{FS_2}, the Fourier transform of $iG^{-+}(t)$ and $iG^{+-}(t)$ by \eq{iGmp_t_f} and \eq{iGpm_t_f} can be expressed as 
\begin{align}
    i\tilde{G}^{-+}(\omega) =&\, p S^{\Psi^{\dagger}\Psi}_{\beta}(\omega-\Omega)+ 
 (1-p) S^{\Psi \Psi^{\dagger}}_{\beta}(\omega + \Omega) \;,
\label{FGmp}
\\
i\tilde{G}^{+-}(\omega) =&\, e^{-\beta \omega} \Big[ p e^{-\beta \Omega} S^{\Psi^{\dagger}\Psi}_{\beta}(\omega + \Omega)+ (1-p) e^{\beta \Omega} \times \nonumber\\
&\,S^{\Psi \Psi^{\dagger}}_{\beta}(\omega - \Omega)  \Big]\;.
\label{FGpm}
\end{align}

We see that $S^{\Psi^{\dagger}\Psi}_{\beta}(\omega)$ can be understood as the quasiparticle spectral density, and $S^{\Psi \Psi^{\dagger}}_{\beta}(\omega)$  as the quasihole spectral density, and they should obey the following KMS condition
\begin{align}\label{f_KMS}
    \begin{cases}
        S^{\Psi^{\dagger}\Psi}_{\beta}(\omega) = n_{\rm FD}(\omega) S^{\Psi}(\omega), \\
        S^{\Psi \Psi^{\dagger}}_{\beta}(\omega) = \big(1-n_{\rm FD}(\omega) \big) S^{\Psi}(\omega) = e^{\beta \omega} n_{\rm FD}(\omega)  S^{\Psi}(\omega)  
    \end{cases}
\end{align}

with
\be
n_{\rm FD}(\omega) ={1\over e^{\beta \omega} + 1}\;.
\ee

Using \eq{FGmp}, \eq{FGpm} and \eq{f_KMS}, we can have
\begin{align}
    &i\tilde{G}^{-+}(\omega)\nonumber\\
    =&\,   p n_{\rm FD}(\omega - \Omega) S^{\Psi}(\omega-\Omega) + (1-p) e^{\beta (\omega +\Omega)} n_{\rm FD}(\omega + \Omega)\nonumber\\
    &\times S^{\Psi}(\omega + \Omega) \;,
\label{FGmp_1}
\\
&i\tilde{G}^{+-}(\omega)\nonumber\\
=&\, p e^{-\beta (\omega +\Omega)} n_{\rm FD}(\omega + \Omega) S^{\Psi}(\omega + \Omega)+(1-p) n_{\rm FD}(\omega - \Omega)\nonumber\\
&\times S^{\Psi}(\omega - \Omega) \;.
\label{FGpm_1}
\end{align}

and thus
\begin{align}
    \label{main_2}
&i\tilde{G}^{-+}(\omega) \pm i\tilde{G}^{+-}(\omega) \nonumber\\
=&\, \big[p \pm (1-p) \big]\; n_{\rm FD}(\omega - \Omega) S^{\Psi}(\omega-\Omega) + \big[(1- p) e^{\beta (\omega + \Omega)} \pm \nonumber\\
&\, p e^{-\beta (\omega +\Omega)}\big]\; n_{\rm FD}(\omega + \Omega) S^{\Psi}(\omega + \Omega) \;. 
\end{align}

This is quite different from \eq{main_1} for the spin coupling. Using \eq{FGmp_1}, \eq{FGpm_1} and \eq{main_2}, we can obtain $\chi^{(2)}(v)$, $P^{(2)}(W)$ and $\overline{W}^{(2)}$ by \eq{chiv_omega_1_m}, \eq{p_W_1_m}, \eq{m_p0} and \eq{W_ext_f}, respectively. 
Analogous with \eq{chi_2_sq}, using \eq{chi_beta2_G} we obtain
\begin{align}
    &\chi^{(2)}(i\beta)\nonumber\\
=&\, 1-{1\over 2} \int_{\lambda} \; \big(1-e^{-\beta \omega} \big) \; \Big\{ 
\big[ p - (1-p) e^{\beta \omega} \big] n_{\rm FD}(\omega - \Omega)  S^{\Psi}(\omega-\Omega)\nonumber\\
&\,  + \big[ (1-p) e^{\beta(\omega +\Omega)} - p e^{-\beta \Omega} \big]  n_{\rm FD}(\omega + \Omega) S^{\Psi}(\omega + \Omega)  \Big\}\;.
\end{align}

\subsection{Work statistics with topological qubit coupling to thermal bath}

On the other hand, when considering the topological qubit, the corresponding real-time Green functions appearing in \eq{chiv_omega_1_m} should be different from the ones given by \eq{Gmp_FQ_t} and \eq{Gpm_FQ_t} due to imposing the causality condition \eq{cluster_p}. Denoting the corresponding Green functions for the topological qubit by $iG^{-+}_{\rm TQ}(t)$ and $iG^{+-}_{\rm TQ}(t)$, we can obtain them by dropping the terms involving  $\langle O_1(t) O_2(t') \rangle_{\beta} - \langle O_2(t) O_1(t') \rangle_{\beta}$ in \eq{Gmp_FQ_t} and \eq{Gpm_FQ_t}, and the results are 
\begin{align}
    iG^{-+}_{\rm TQ}(t) =&\, {1\over 2} iG^{-+}_{\gamma,11}(t) \Big(  \langle  \Psi^{\dagger}(t) \Psi(0) \rangle_{\beta} + \langle  \Psi(t) \Psi^{\dagger}(0) \rangle_{\beta} \Big)\;,
\\
iG^{+-}_{\rm TQ}(t) =&\, {1\over 2} iG^{-+}_{\gamma,11}(-t) \Big(  \langle  \Psi^{\dagger}(t-i\beta) \Psi(0) \rangle_{\beta} + \nonumber\\
&\,\langle  \Psi(t-i\beta) \Psi^{\dagger}(0) \rangle_{\beta} \Big)\;.
\end{align}

Their Fourier transforms are then
\begin{align}
    \label{Gmp_TQ_1}
i\tilde{G}^{-+}_{\rm TQ}(\omega) =&\, p S^{\gamma}(\omega-\Omega) + (1-p) S^{\gamma}(\omega + \Omega)\;, 
\\
i\tilde{G}^{+-}_{\rm TQ}(\omega) =&\, e^{-\beta \omega} \Big[ p e^{-\beta \Omega} S^{\gamma}(\omega + \Omega) + (1-p) e^{\beta \Omega}\nonumber\\
&\, \times S^{\gamma}(\omega - \Omega) \Big] \label{Gpm_TQ_1}
\end{align}
where we have defined the spectral density of MZM by
\be\label{S_gamma}
S^{\gamma}(\omega) = {1\over 2} \Big(S^{\Psi \Psi^{\dagger}}_{\beta}(\omega) + S^{\Psi^{\dagger}\Psi}_{\beta}(\omega)  \Big) = {1\over 2} S^{\Psi}(\omega) \;.
\ee
From \eq{Gmp_TQ_1}, \eq{Gpm_TQ_1} and \eq{S_gamma}, we have
 \begin{align}
     \label{main_3}
&i\tilde{G}^{-+}_{\rm TQ}(\omega) \pm i\tilde{G}^{+-}_{\rm TQ}(\omega) \nonumber\\
=&\, {1\over 2} \big[ 1-p \pm p^{-\beta(\omega+\Omega)}\big]\; S^{\Psi}(\omega + \Omega) + {1\over 2} \big[p \pm (1-p) e^{-\beta (\omega-\Omega)} \big]\; \nonumber\\
&\, \times S^{\Psi}(\omega - \Omega) 
 \end{align}
This form is quite different from \eq{main_2} but is similar to \eq{main_1} with $S_{\beta}^O(\omega)$ replaced by ${1\over 2} S^{\Psi}(\omega)$.  Using \eq{main_3}, we can obtain $\chi^{(2)}(v)$, $P^{(2)}(W)$ and $\overline{W}^{(2)}$ by \eq{chiv_omega_1_m}, \eq{p_W_1_m}, \eq{m_p0} and \eq{W_ext_f}, respectively. 
Analogous with \eq{chi_2_sq}, using \eq{chi_beta2_G}, we obtain
\begin{align}
    \chi^{(2)}(i\beta)=&\,1-{1\over 4} \int_{\lambda} \big(1-e^{-\beta \omega} \big) \; \Big\{\big[ 1- p - p e^{-\beta \Omega} \big]  S^{\Psi}(\omega+ \Omega)\nonumber\\
    &\,   + \big[p - (1-p) e^{\beta \Omega} \big] S^{\Psi}(\omega - \Omega)  \Big\}\;.
\end{align}

\onecolumngrid

\section{Numerical result supplement of work statistics}\label{app_E}

To give more thorough information on the work statistics based on our EFT approach for the Ohmic-type thermal bath with and without qubit coupling, we present in this section more pedagogical numerical results about the work distribution function (WDF) $P^{(2)}(W)$ and work extraction $W^{(2)}_{\text{ext}}$. These plots also include the ones mentioned but not presented in the main text.

The setup of our EFT approach to the work statistics of a cyclic process via qubit coupling to the thermal bath comprises seven parameters: three about the thermal bath, i.e., $\alpha$, $\beta$, and $L_c$, two about control of the cyclic process, i.e., $\lambda_0$ and $T_{\rm int}$, and two about the qubit system, i.e., $p$ ad $\Omega$. To ensure the validity of EFT, the cyclic process should be adiabatic, i.e., $T_{\rm int}>L_c$. For simplicity, we fix $L_c=1$ and $T_{\rm int}=100$. Besides, we restrict the consideration to the perturbative expansion of the evolution operator by setting $\lambda_0=0.01$. Thus, the plots presented in this paper for $P^{(2)}$ and $W^{(2)}_{\rm ext}$ will depend on the chosen values of $\alpha$, $\beta$, $p$, and $\Omega$.

\subsection{Work Distribution Function $P^{(2)}(W)$}\label{app_E1}
\noindent
    In this subsection, we present more numerical plots for $P^{(2)}(W)$ for the pure thermal baths without or with three types of qubit couplings for various values of $\alpha$, $\beta$, $p$, and $\Omega$ to highlight some interesting points about the modal structure and asymmetry of $P^{(2)}(W)$. For simplicity, we neglect $p_0 \delta(W)$ of $P^{(2)}(W)$  in the following plots.

\begin{enumerate}[leftmargin=0pt]

    \item We first compare $P^{(2)}(W)$ with different values of $\Omega$ with other parameters fixed. The plots for $\Omega=1, 5$ are given in Fig. \ref{prl_pw_sup1}. It shows that $P^{(2)}(W)$ are almost symmetric about $W=0$. This  contrasts with Fig. \ref{prl_pw}(c) in the main text for $\Omega=0.05$, which show asymmetric $P^{(2)}(W)$ about $W=0$. This implies that $P^{(2)}(W)$ tends to be asymmetric about $W=0$ as $\Omega$ decreases.

\begin{figure}[hbt!]
	\centering
    \includegraphics[width=0.8\linewidth]{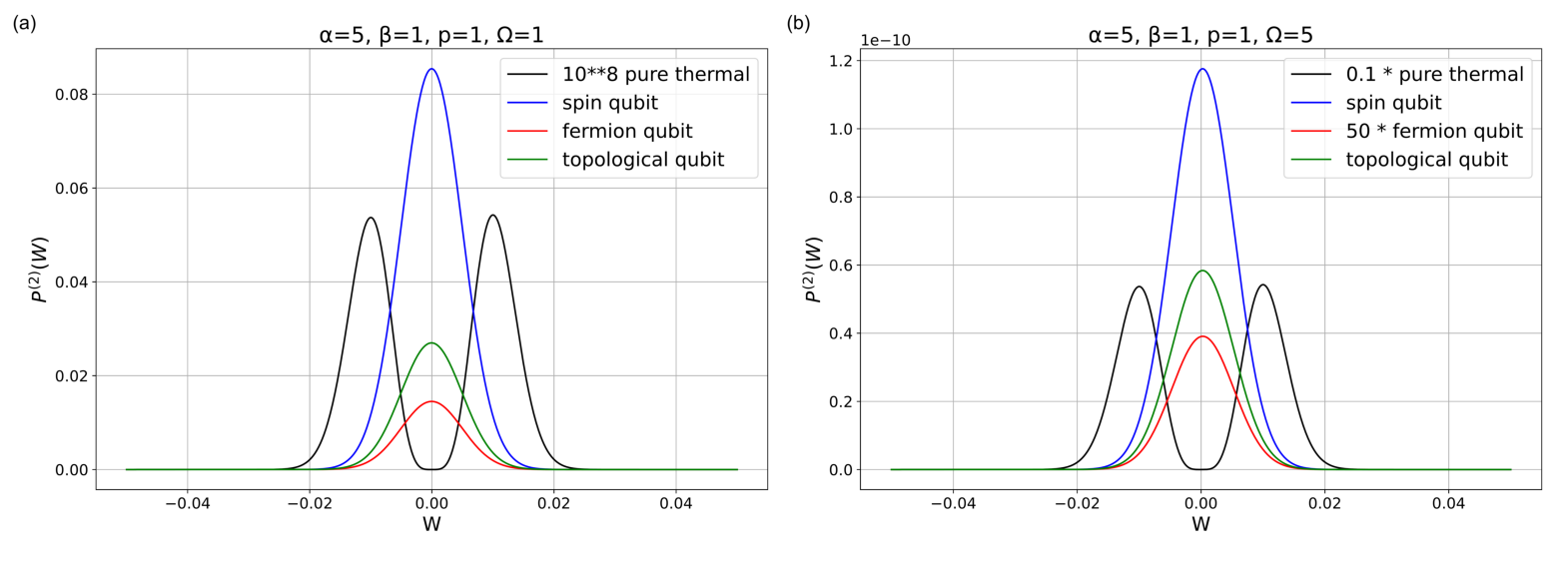}
    \caption{Compared to Fig. \ref{prl_pw}(c) in the main text, it implies that $P^{(2)}(W)$ tends to be asymmetric about $W=0$ as $\Omega$ decreases.
    }
    \label{prl_pw_sup1}
\end{figure}

    \item We show examples of unimodal $P^{(2)}(W)$'s in Fig. \ref{prl_pw_sup2} with or without qubit coupling, all with $\alpha\le 1$. This contrasts with the bimodal ones for the pure thermal bath with $\alpha>1$ as shown in Fig. \ref{prl_pw}(b) of the main text.  These plots help to conclude that the $P^{(2)}(W)$ is bimodal only for $\alpha>1$ without qubit coupling.
    
\begin{figure}[hbt!]
	\centering
    \includegraphics[width=0.8\linewidth]{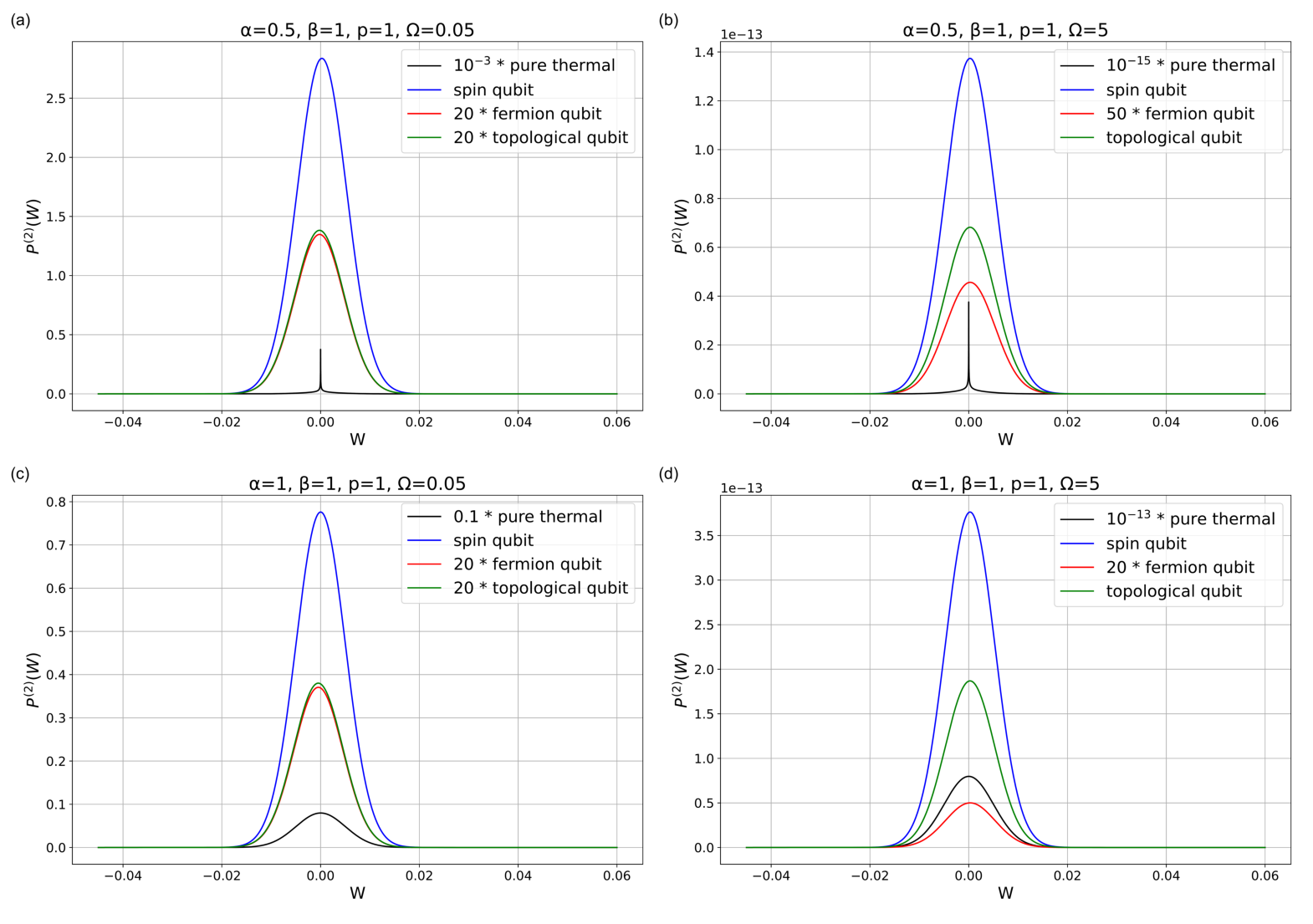}
    \caption{Examples of unimodal $P^{(2)}(W)$'s.
    }
    \label{prl_pw_sup2}
\end{figure}

    \item We show in Fig. \ref{prl_pw_sup3} that the asymmetry of $P^{(2)}(W)$ about $W=0$ changes its bias as $\beta$ changes.

\begin{figure}[hbt!]
	\centering
    \includegraphics[width=0.8\linewidth]{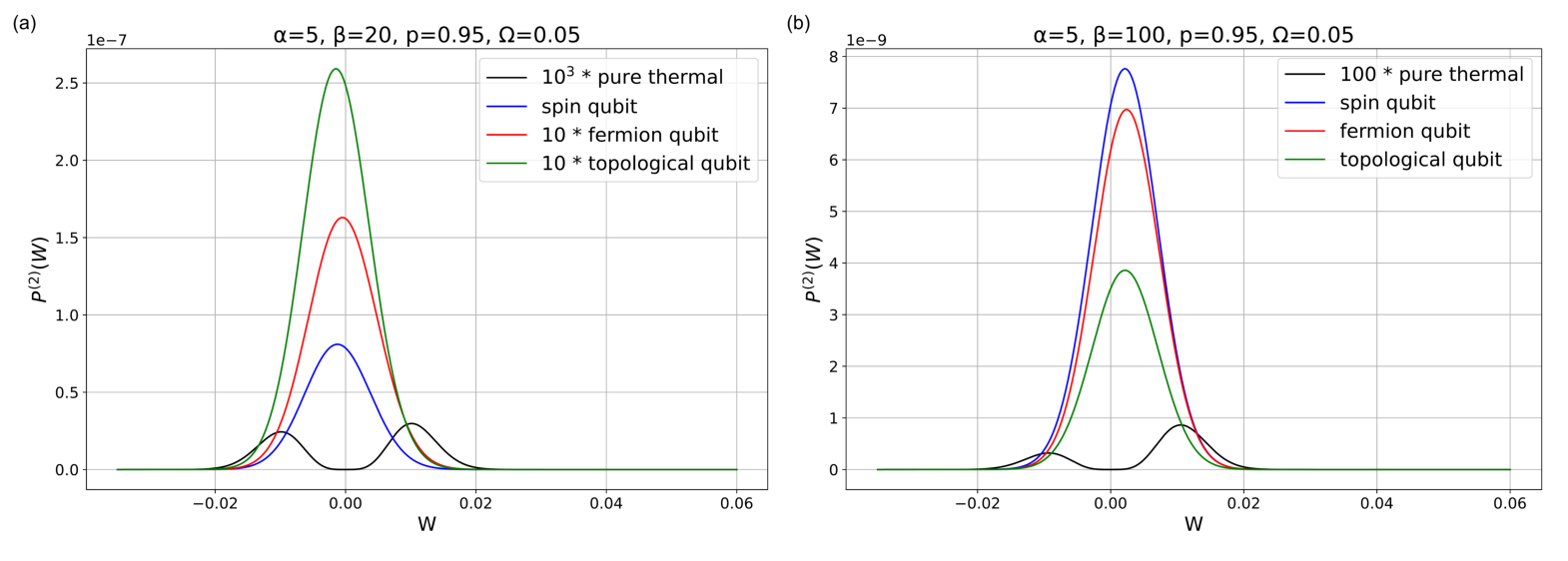}
    \caption{Two asymmetric $P^{(2)}$'s with qubit coupling correspond respectively to $W^{(2)}_{\rm ext}< 0$ for $\beta=20$ (left) and $W^{(2)}_{\rm ext} > 0$ for $\beta=100$ (right).
    }
   \label{prl_pw_sup3}
\end{figure}
    
\end{enumerate}

\begin{figure}[hbt!]
	\centering
    \includegraphics[width=0.8\linewidth]{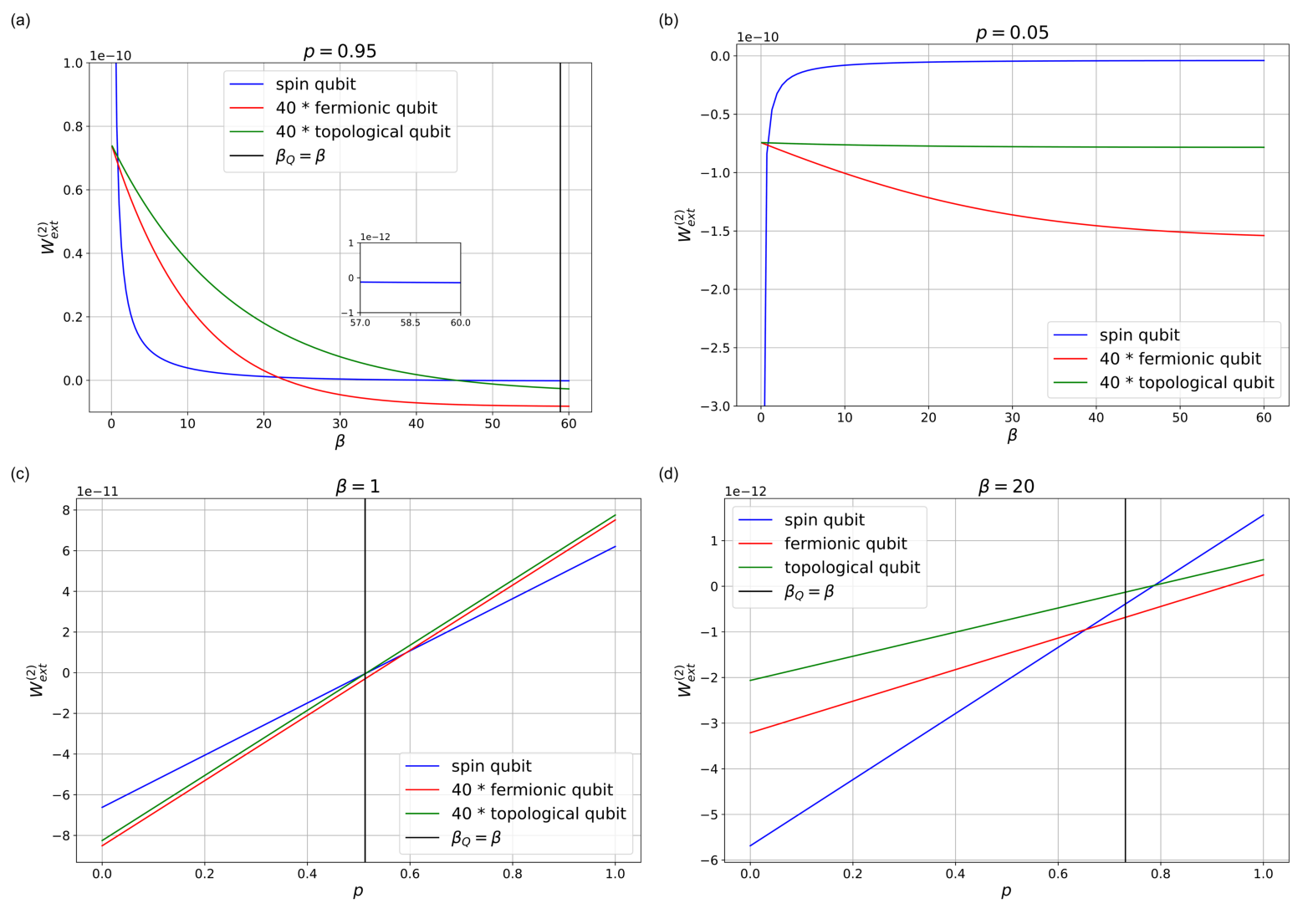}
    \caption{$W^{(2)}_{\rm ext}$ vs $\beta$ for (a) $p=0.95$ and (b) $p=0.05$, and $W^{(2)}_{\rm ext}$ vs $p$ for (c) $\beta=1$ and (d) $\beta=20$ with fixed $alpha=5$ and $\Omega=0.05$. The black lines in the subfigures denote $\beta_Q=\beta$ regime.  
    }
    \label{prl_w_ext_sup1}
    \end{figure}

\begin{figure}[hbt!]
	\centering
    \includegraphics[width=0.8\linewidth]{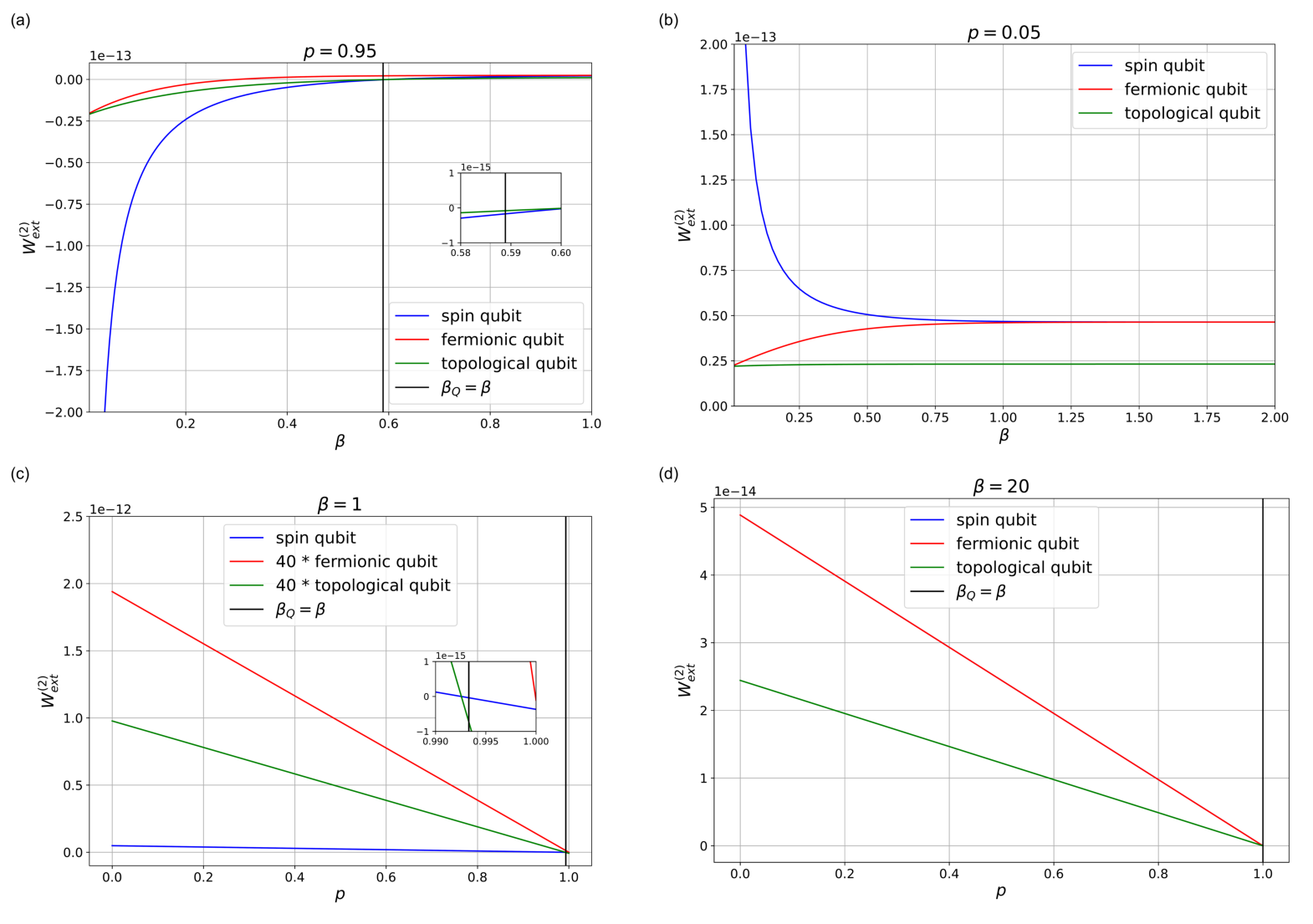}
    \caption{Simialr plots to Fig. \ref{prl_w_ext_sup1} but with $\Omega=5$. In subfigure (d), the lines of spin and fermion qubit coupling almost overlap. 
    }
    \label{prl_w_ext_sup2}
\end{figure}

\subsection{Work Extraction $W_{\text{ext}}^{(2)}$} \label{app_E2}

Though the density plots of $W_{\rm ext}^{(2)}$ can give a more complete picture, it is not easy to compare different types of qubit coupling. Alternatively, here we present the 1D plots of $W_{\rm ext}^{(2)}$ on only one parameter but with others fixed for all three qubit-coupling cases in a single figure. In this way, we can see how the patterns of $W_{\rm ext}^{(2)}$ depend on the types of qubit couplings. 

In Fig. \ref{prl_w_ext_sup1} (Fig. \ref{prl_w_ext_sup2}), we fix $\Omega=0.05$ ($\Omega=5$) and show $W_{\rm ext}^{(2)}$ vs $\beta$ in subfigures (a) and (b) for $p=0.95$ and $p=0.05$, respectively; and $W_{\rm ext}^{(2)}$ vs $\beta$ in subfigures (c) and (d) for $\beta=1$ and $\beta=20$, respectively. 

These plots imply the following: (i) the $W_{\rm ext}^{(2)}$ plots in Fig. \ref{prl_w_ext_sup1} for $\Omega=0.05$ and Fig. \ref{prl_w_ext_sup2} for $\Omega=5$ show completely opposite behaviors; (ii) $W_{\rm ext}^{(2)}$ for fixed $p$ saturates at large $\beta$; (iii) $W_{\rm ext}^{(2)}$ shows strict linear dependence on $p$, as can be easily understood; (iv)  $W^{(2)}_{\rm ext}>0$ occurs almost for all $p$ for large enough $\Omega$ and $\beta$. This is consistent with Fig. \ref{prl_w_ext} of the main text.

\FloatBarrier

\twocolumngrid

\bibliography{work_ref.bib} 

\end{document}